\documentclass[12pt,a4paper]{article}
\usepackage[latin1]{inputenc}
\usepackage{amssymb,amsmath,mathrsfs,bm}
\usepackage{comment}

 \usepackage{mdframed}
 \usepackage{graphicx}
 \usepackage{setspace}
 \usepackage{a4wide}
\usepackage{tcolorbox}
\usepackage{cite}
\usepackage{hyperref}

\usepackage{tikz}
\usepackage{color}

\makeatletter 
\@addtoreset{equation}{section}
\makeatother

\hypersetup{
  linktocpage,
  colorlinks  = true, 
  urlcolor    = blue, 
 linkcolor   = red, 
  citecolor   = blue 
}

\def \be  {\begin{equation}}
\def \ee  {\end{equation}}
\def \ba  {\begin{eqnarray}}
\def \ea  {\end{eqnarray}}

\def \Tr {\mathop{\rm Tr}\nolimits}

\usepackage{titlesec}

\titleformat*{\section}{\large\bfseries}

\begin{document}

\usetikzlibrary{arrows}
\thispagestyle{empty}

~\\[-2.25cm]


\begin{center}
{\Large\bf
%
{\Large The Veneziano amplitude in AdS$_5 \times$S$^3$ from an 8-dimensional effective action}
}\\
\vskip 1.25truecm
	{\bf R. Glew$^a$ and M. Santagata$^b$ \\
	}
		\vskip 0.4truecm
 
	{\it
		$^a$Department of Physics, Astronomy and Mathematics, University of Hertfordshire, Hatfield, Hertfordhshire, AL10 9AB, United Kingdom. \\
		$^b$Department of Physics, National Taiwan University, Taipei 10617, Taiwan. \\
		\vskip .2truecm                        }
	\vskip .2truecm

\end{center}

\textit{E-mail:}\,\, \href{mailto:michelesa@ntu.edu.tw}{{\tt michelesa@ntu.edu.tw}}, \,\, \href{mailto:r.glew@herts.ac.uk}{{\tt r.glew@herts.ac.uk}}

\vskip 1.25truecm 

\vskip 1.25truecm 

\centerline{\bf Abstract} 
\vskip .4truecm
We study four-point functions of arbitrary half-BPS operators in a 4-dimensional $\mathcal{N}=2$ SCFT with flavour group $SO(8)$ at genus-zero and strong 't Hooft coupling, corresponding - via AdS/CFT - to the ($\alpha'$ expansion of the) Veneziano amplitude on an AdS$_5 \times$S$^3$ background. We adapt a procedure first proposed by Abl, Heslop and Lipstein in the context of AdS$_5 \times$S$^5$, and postulate the existence of an effective action in terms of an $8$-dimensional scalar field valued in the adjoint of the flavour group. 
The various Kaluza-Klein correlators can then be computed by uplifting the standard AdS/CFT prescription to the full product geometry with AdS bulk-to-boundary propagators and Witten diagrams replaced by suitable AdS$_5 \times$S$^3$ versions.
After elucidating the main features of the procedure, valid at all orders in $\alpha'$, we show explicit results up to order $\alpha'^{5}$.
The results provide further evidence of a novel relation between AdS$\times$S and flat amplitudes - which made its first appearance in $\mathcal{N}=4$ SYM - that is perhaps the most natural extension of the well known flat-space limit proposed by Penedones to cases where AdS and S have the same radius.

\noindent

\vskip 1truecm 


\newpage
\setcounter{page}{1}\setcounter{footnote}{0}
\tableofcontents

\newpage


\section{Introduction and summary of results}\label{intro_sec}


In recent years major progress has been made in the computation of holographic correlators, i.e. correlators in CFT's that admit a gravity dual. It is reasonable to hope that understanding the structure of these observables will shed light on the properties of quantum gravity and its UV completion.

In the well studied case of $\mathcal{N}=4$ SYM, dual to string theory on AdS$_5 \times$S$^5$ \cite{Maldacena:1997re}, a number of different bootstrap approaches have led to a systematic understanding of the structure of four-point functions of half-BPS operators at various order in $1/N$ and $1/\lambda$\cite{Rastelli:2017udc,Rastelli:2016nze,Alday:2018pdi,Alday:2018kkw,Alday:2019nin,Aprile:2017qoy,Aprile:2017bgs,Aprile:2019rep,Drummond:2019odu,Drummond:2019hel,Goncalves:2014ffa,Drummond:2020dwr,Aprile:2020mus,Bissi:2020wtv,Bissi:2020woe,Aprile:2020luw,Drummond:2020uni,Aprile:2022tzr,Huang:2021xws,Drummond:2022dxw,Alday:2022uxp,Alday:2022xwz}\footnote{See \cite{Heslop:2022xgp,Bissi:2022mrs} for recent reviews on the subject.} .
Despite the apparent complexity of the intermediate steps, the end results are found to be tremendously simple. This is no coincidence: amazingly, it turns out that $\mathcal{N}=4$ SYM enjoys a hidden $10d$ conformal symmetry in the supergravity limit\cite{Caron-Huot:2018kta}\footnote{The symmetry has also been observed at weak coupling in $\mathcal{N}=4$ SYM \cite{Caron-Huot:2021usw} and, recently, extended to higher components of the stress-tensor multiplet \cite{Caron-Huot:2023wdh}.}.
This symmetry allows a repackaging of all four-point correlators into a single $10d$ object transforming as a four-point function of weight $4$ scalars.

Away from supergravity, when $\alpha'$ corrections are switched on, the hidden conformal symmetry is broken; however, string corrections are still found to obey a $10d$ principle \cite{Drummond:2020dwr,Aprile:2020mus,Abl:2020dbx}.
Specifically, the authors of \cite{Abl:2020dbx} postulate the existence of a 10-dimensional effective action in terms of a single scalar such that, the correlators for \emph{arbitrary} Kaluza-Klein modes computed out of the action via Witten diagrams, perfectly agree with all known results in the literature \cite{Goncalves:2014ffa,Drummond:2019odu,Drummond:2020dwr,Aprile:2020mus}.
The underlying idea of the computation is to mimic the standard AdS/CFT procedure, with the difference that standard AdS Witten diagrams are replaced by generalised AdS$\times$S versions. 
The method leaves a handful of free ambiguities at each order in $\alpha'$ which can be interpreted as an effect of the curvature of the background.

A natural question to ask is whether this approach can be useful in other physical theories of interest\footnote{See also \cite{Abl:2021mxo}, where this approach has been used to compute higher-derivative corrections in AdS$_2\times$S$^2$. Another theory where we expect these methods to be effective are the AdS$_3\times$S$^3$ correlators of \cite{Giusto:2019pxc,Rastelli:2019gtj,Giusto:2020neo}.}.
A promising candidate in this respect is a recently studied 4-dimensional $\mathcal{N}=2$ theory with global group $SO(8)$. This is the theory  we consider in this paper.
This SCFT describes the dynamics of $N$ D3-branes near a F-theory 7-brane singularity \cite{Sen:1996vd,Fayyazuddin:1998fb,Aharony:1998xz}.
In this model, four-point functions of chiral primary operators correspond to the scattering of four supergluons in a AdS$_5 \times$S$^3$ bulk with the higher derivative corrections representing the string completion of the field-theory amplitude, which is an AdS$_5 \times$S$^3$ version of the Veneziano amplitude. These correlators have already been studied in the field-theory limit in a series of papers both at tree-level, one and two loops \cite{Alday:2021odx,Alday:2021ajh,Huang:2023oxf,Drummond:2022dxd} but little is known for the $\alpha'$ corrections.
However, given the many similarities shared by $\mathcal{N}=4$ SYM and the SCFT considered in this paper, in particular the existence of a \emph{reduced} correlator and the fact that this can be re-organized into a single $8$-dimensional object\cite{Alday:2021odx},  we expect the methods of \cite{Abl:2020dbx} to naturally generalise to this background.

This motivates us to conjecture the existence of an $8d$ action written in terms of a scalar field valued in the adjoint of $SO(8)$, which encodes tree-level correlators of super gluons with \emph{arbitrary} Kaluza-Klein levels up to a small number of ambiguities.
The main result of this paper is to spell out the effectiveness of the method and compute explicitly these correlators at the first few orders in $\alpha'$.

The starting point is to conceive the ($\alpha'$ expansion of the) Veneziano amplitude as arising from an effective potential 
\begin{equation}
\begin{split}
V_{\text{open}}=& \frac{1}{8}\frac{\pi^2}{6} \Tr[\phi^4]\alpha'^2 +\frac{1}{2} \zeta_3 \Tr \left[ (\partial_\mu \phi) (\partial_\mu \phi)  \phi \phi \right]\alpha'^3 + \\
&+ \frac{1}{2} \frac{\pi^4 }{720} \Tr \left[14(\partial_\mu \partial_\nu \phi) (\partial_\mu \partial_\nu \phi)\phi^2+(\partial_\mu \partial_\nu \phi)\phi (\partial_\mu \partial_\nu \phi)\phi\right]\alpha'^4-\\
& +\frac{1}{3}\Tr \left[2 \left(\frac{1}{6}\zeta_3 \pi^2+\zeta_5\right)(\partial_\mu \partial_\nu \partial_\rho \phi) (\partial_\mu \partial_\nu  \partial_\rho \phi)\phi^2+ \right.\\
& +\left. \left(\frac{1}{6}\zeta_3 \pi^2-2\zeta_5\right)(\partial_\mu \partial_\nu \partial_\rho \phi) \phi(\partial_\mu \partial_\nu  \partial_\rho \phi)\phi\right] \alpha'^5+ \cdots,
\end{split}
\end{equation} 
where the constants are fixed by requiring that the four-point amplitude in momentum space at a given order in $\alpha'$ matches the corresponding term in the low-energy expansion of the flat-space Veneziano amplitude. 
The idea is then to uplift this potential to AdS$_5 \times$S$^3$ by replacing flat derivatives with AdS$_5 \times$S$^3$ covariant versions. 
The various Kaluza-Klein correlators can eventually be computed by using a generalisation of contact Witten diagrams which takes into account the compact space.
Nicely, we find that the end results are simple functions of AdS and S variables. 
In particular, we notice that they can be written in a very compact form in terms of a pre-amplitude which made its first appearance in $\mathcal{N}=4$ SYM \cite{Aprile:2020mus}. 
Interestingly, this said pre-amplitude shares a strong similarity with its flat-space counterpart and it is related to the latter via a double integral transform, which is perhaps the most natural generalisation of the integral transform defining the flat-space limit conjectured by Penedones \cite{Penedones:2010ue} and later proved in \cite{Fitzpatrick:2011hu}. 

To give a flavour for the remarkable simplicity of the results, the order $\alpha'^3$ correlator for \emph{arbitrary} Kaluza-Klein correlators reads
\begin{align}
& \tilde{\mathcal{M}}_{p_1p_2p_3p_4}(1234) \big|_{\alpha'^3}= \zeta_3 \left( S + T  + a_1 \right)
\end{align}
where $\tilde{\mathcal{M}}_{p_1p_2p_3p_4}(1234) $ is a colour-ordered Mellin \emph{pre-amplitude}, $S,T$ are suitable AdS$_5 \times$S$^3$ variables, and $a_1$ is the only ambiguity left at this order. While the notation will be explained in full in the main body, we can already appreciate the manifest similarity with the flat-space Veneziano amplitude which at this order reads
\begin{align}
& \mathcal{V}_{\text{open}} (1234)\big|_{\alpha'^3} = \zeta_3 \left( s+t \right).
\end{align}
where $s,t$ are the Mandelstam variables.
The compactness of these results seems to point out the existence of an underlying structure yet to be discovered.

The remainder of the paper is organised as follows. In section \ref{sec:gen} we describe the general set-up for the object of study here: four-point correlation functions of half-BPS operators dual to supergluon amplitudes on AdS$_5 \times$S$^3$. In section \ref{sec:venac} we review the structure of the Veneziano amplitude in flat space as well as the procedure outlined in \cite{Abl:2020dbx} for uplifting to AdS via viewing the correlator as arising from an $8d$ scalar effective action.
In section \ref{sec:expliv} we give explicit results for the $\alpha'$-corrected correlators up to order $\alpha'^5$. In the last part of this section we clarify the relation between the flat-space limit known in literature and some possible generalisations.
Finally, in Section  \ref{sec:conc} we comment on possible future directions. 

{\bf Note added}: Whilst completing our paper, we were informed by the authors of \cite{Behan:2023fqq} of their work on a similar topic.
We thank them for coordinating the release on the arXiv.

\section{Generalities}
\label{sec:gen}
In this paper we are interested in the scattering of super gluons in an AdS$_5 \times$S$^3$ background.
As mentioned in the introduction, the dual SCFT is a four-dimensional $\mathcal{N}=2$ theory with a certain global group, which plays the role of a gauge group in the bulk. This SCFT arises as the worldvolume theory of D3-branes moving near F-theory 7-branes singularities \cite{Sen:1996vd,Fayyazuddin:1998fb,Aharony:1998xz}.
In the case we are interested in, the 7-branes correspond to a $\mathbb{Z}_2$ orientifold point, with the low-energy dynamics of $N$ D3-branes described by a $USp(2N)$ $\mathcal{N}=2$ gauge theory with an additional $SO(8)$ global symmetry group.
Let us clarify that it is in principle possible to consider different orbifolds, which give rise to SCFTs with different global symmetries \cite{Fayyazuddin:1998fb,Aharony:1998xz}. Although, as we will see, the computations considered in this paper could be performed without making any reference to the gauge group, all other constructions do not have a perturbative formulation\footnote{Another possibility would be to consider a small number of flavour D7-branes in a D3 background, following the construction in \cite{Karch:2002sh}. In the limit $N_F \ll  N$, this gives rise to a $\mathcal{N}=2$ SCFT with flavour group $SU(N_F)$, dual to SYM on AdS$_5 \times$S$^3$ background. As mentioned earlier, our approach is agnostic about the flavour group, thus it can in principle be also applied to this theory. We thank an anonymous referee for pointing this out.}. In other words, the bulk theory does not have an $\alpha'$ parameter (equivalently, from a CFT perspective, there is no marginal coupling) and an $\alpha'$ expansion would evidently be meaningless\footnote{We thank Pietro Ferrero for discussion on this point.}.

The presence of the 7-branes breaks the $SO(6)$ isometry group of S$^5$ to $SU(2)_R\times SU(2)_L \times U(1)_R$. From the point of view of the dual $\mathcal{N}=2$ SCFT, $SU(2)_R\times U(1)_R$ becomes the R-symmetry group and $SU(2)_L$ is an additional global group.

We are interested in the (scalar component of) the $\mathcal{N}=1$ vector multiplet and its Kaluza-Klein tower which are dual to half-BPS scalar operators of the form $\mathcal{O}_p^{I a_1 \ldots a_p; \bar{a}_1\ldots \bar{a}_{p-2}}$. These operators are chargeless under $U(1)_R$, they transform under the spin $\frac{p}{2}$ of $SU(2)_R$, spin $\frac{p}{2}-1$ of $SU(2)_L$ and in the adjoint of $SO(8)$.
Here $I$ is the colour index, $p$ is the scaling dimension of the operator, $a_1,\ldots, a_p$ are symmetrised $SU(2)_R$ R-symmetry indices and similarly $\bar{a}_i$ are indices of an additional $SU(2)_L$ flavour group; these last two groups realise the isometry group of the sphere S$^3$. 
A convenient way to deal with the $SU(2)_R \times SU(2)_L$ indices is by contracting them with auxiliary bosonic two-component vectors $\eta$ and $\bar{\eta}$:
\begin{equation}
\label{contractedO}
\mathcal{O}_p^{I} \equiv \mathcal{O}_p^{I;a_1 \ldots a_p; \bar{a}_1\ldots \bar{a}_{p-2}}\eta_{a_1} \ldots \eta_{a_p} \bar{\eta}_{\bar{a}_1} \ldots \bar{\eta}_{\bar{a}_{p-2}}\,.
\end{equation}
We will denote the four-point function of half-BPS operators by
\begin{equation}
G_{\vec{p}}^{I_1 I_2 I_3 I_4}(x_i,\eta_i, \bar{\eta}_i)\equiv \langle \mathcal{O}_{p_1}^{I_1} \mathcal{O}_{p_2}^{I_2} \mathcal{O}_{p_3}^{I_3} \mathcal{O}_{p_4}^{I_4} \rangle.
\end{equation} 
Note that the correlator is a function of $x_i,\eta_i, \bar{\eta}_i$ and the charges $p_i$.
In particular, due to the definition \eqref{contractedO}, it is a polynomial in the variables $\eta_i, \bar{\eta}_i$, whose degree is dictated by the external charges $p_i$.
The variables $\eta_i,\bar{\eta}_i$ are contracted via $\langle \eta_i \eta_j\rangle = \eta_{ia} \eta_{jb} \epsilon^{ab}$ and $\langle \bar{\eta}_i \bar{\eta}_j\rangle = \bar{\eta}_{i\bar{a}} \bar{\eta}_{j\bar{b}} \epsilon^{\bar{a}\bar{b}}$.

$G_{\vec{p}}^{I_1 I_2 I_3 I_4}$ is subject to the superconformal Ward identities \cite{Nirschl:2004pa}
\be
G_{{\vec{p}}}^{I_1 I_2 I_3 I_4} = G_{\text{free},\vec{p}}^{I_1 I_2 I_3 I_4}  +\, I \, G_{\text{int},{\vec{p}}}^{I_1 I_2 I_3 I_4}\, ,
\ee
where the kinematic factor $I$ takes the following form:
\begin{equation}
\label{Idefin}
I= x_{13}^2 x_{24}^2 \langle \eta_1 \eta_3 \rangle^2 \langle \eta_2 \eta_4 \rangle^2 (x-y) (\bar{x}-y),
\end{equation}
and we have defined 
\begin{equation}
x \bar{x}= \frac{x_{12}^2x_{34}^2}{x_{13}^2 x_{24}^2}, \qquad y= \frac{\langle \eta_1 \eta_2 \rangle \langle \eta_3 \eta_4 \rangle}{\langle \eta_1 \eta_3 \rangle\langle \eta_1 \eta_3 \rangle}.
\end{equation}
The non-trivial function left over, denoted by $G_{\text{int},{\vec{p}}}^{I_1 I_2 I_3 I_4}$, has two units of conformal/internal weights less than $G_{\vec{p}}^{I_1 I_2 I_3 I_4}$.
As a consequence, $G_{\text{int},{\vec{p}}}^{I_1 I_2 I_3 I_4}$ is a polynomial of the \emph{same} degree in $\eta,\bar{\eta}$. 

The function $G_{\text{int},{\vec{p}}}^{I_1 I_2 I_3 I_4}$ admits a genus ($\frac{1}{N}$) expansion. In this paper we are interested in the $\mathcal{O}(\frac{1}{N})$ order, which corresponds to a bulk tree-level amplitude.
This can be further expanded in the (square of the) string length $\alpha'$.\footnote{This a strong-coupling expansion in the CFT, with the Regge slope related to the Yang-Mills coupling via $\frac{R^4}{\alpha'^2}=g_{\text{YM}}^2 N$.} Our notation for the $\alpha'$ expansion of the tree-level amplitude reads
\begin{equation}
\label{positionspaceexp}
G_{\text{tree},\vec{p}}^{I_1 I_2 I_3 I_4}=G_{\text{YM},\vec{p}}^{I_1 I_2 I_3 I_4}+ G_{0,\vec{p}}^{I_1 I_2 I_3 I_4}\alpha'^2+G_{1,\vec{p}}^{I_1 I_2 I_3 I_4}\alpha'^3+G_{2,\vec{p}}^{I_1 I_2 I_3 I_4}\alpha'^4+G_{3,\vec{p}}^{I_1 I_2 I_3 I_4}\alpha'^5+\cdots
\end{equation}
where $G_{\text{YM},\vec{p}}^{I_1 I_2 I_3 I_4}$ is the field-theory amplitude, first computed in \cite{Alday:2021odx}.
Note that, as explained in \cite{Alday:2021odx}, the graviton exchange is $1/N$-suppressed and can be neglected at this order. Thus, the only massless fields exchanged at tree-level are the gluons themselves.
The goal of the next sections is to outline a procedure to compute the various $G_{i,\vec{p}}^{I_1 I_2 I_3 I_4}$ order by order in $\alpha'$. We will give explicit results for the first four string corrections, i.e. up to order $\alpha'^5$.

The function $G_{\text{int},\vec{p}}^{I_1 I_2 I_3 I_4}$ in general depends on the spacetime variables $x_i$, the R-symmetry variables $\eta_i$, the internal variables $\bar{\eta}_i$, and the charges $p_i$.
However, it turns out that  $G_{\text{tree},\vec{p}}^{I_1 I_2 I_3 I_4}$ is symmetric under $\eta \leftrightarrow \bar{\eta}$ exchange \cite{Alday:2021odx}. As a consequence, the $SU(2)_L \times SU(2)_R$ variables can be reorganized into $SO(4)$ variables. In other words, $G_{\text{tree},\vec{p}}^{I_1 I_2 I_3 I_4}$ ultimately depends on the charges $p_i$, the spacetime distances $x_{ij}^2$, and the $SO(4)$ distances
\begin{equation}
y_{ij}^2= \langle \eta_i \eta_j \rangle \langle \bar{\eta}_i \bar{\eta}_j \rangle.
\end{equation}
We will make use of this non-trivial fact\footnote{A priori, this is \emph{not} guaranteed. For example, the disconnected correlator, as computed by Wick contractions, it is not symmetric under $\eta \leftrightarrow \bar{\eta}$ exchange \cite{Alday:2021ajh,Drummond:2022dxd}.} in the next section to express $G_{\text{tree},{\vec{p}}}^{I_1 I_2 I_3 I_4}$ in embedding $SO(2,4)\times SO(4)$ coordinates.
Note that, unlike $\mathcal{N}=4$ SYM, the factor $I$ - which is entirely due to superconformal symmetry, and therefore is only a function of $x_i,\eta_i$ - cannot be written in terms of embedding coordinates.

Finally, let us conclude this section by defining the \emph{generator} of all correlators
\begin{equation}
\langle \mathcal{O}^{I_1} \mathcal{O}^{I_2} \mathcal{O}^{I_3}\mathcal{O}^{I_4} \rangle \equiv \sum_{\{p_i\}} G_{\text{tree},\vec{p}}^{I_1 I_2 I_3 I_4}
\end{equation}
where the sum is performed over all charges $p_i=2,\ldots,\infty$.
This is a rather natural object to consider in this formalism. In fact, as we will see, a very convenient aspect of this method is that it automatically collects all KK correlators into a single function.

\section{An $8d$ effective action for the Veneziano amplitude}
\label{sec:venac}
The purpose of this section is to outline a procedure for computing the $\alpha'$ expansion of the Veneziano amplitude in AdS$_5 \times$S$^3$.
As already mentioned, this method was first proposed in the context of AdS$_5 \times$S$^5$ \cite{Abl:2020dbx}, and later applied to higher derivative corrections in AdS$_2\times$S$^2$ \cite{Abl:2021mxo}. Perhaps not surprisingly, we find that a straightforward generalisation of their results leads to a very concrete proposal for tree-level correlators in this background at all orders in $\alpha'$.

We should perhaps stress again that we are not considering the actual supersymmetric AdS$_5 \times$S$^3$ action dual to the CFT we are interested in, but a bosonic version of it, thus \emph{a priori} they do not need to be related.
Intuitively, the reason why we expect this simplified bosonic action to capture the correct supersymmetric 8-dimensional Veneziano amplitude is that the \emph{reduced} correlator behaves like a bosonic object. Moreover, since string corrections have the form of contact interactions, it is reasonable to think that they are somewhat insensitive to superpartners.

Beyond these naive (and arguable) arguments, we will see more concretely that the correlators so computed have the correct flat-space limit \cite{Penedones:2010ue,Fitzpatrick:2011hu} and large $p$ limit \cite{Aprile:2020luw}, providing a first check of the correctness of the results. 
In the conclusions we will comment on other possible independent approaches which could help to prove the effectiveness of the method.

\subsection{The flat-space Veneziano amplitude}
Let us start by recalling the form of the supersymmetric Veneziano amplitude in flat space and some of its properties. 
With a slight abuse of language, with Veneziano amplitude, we will refer, here and after, to the amplitude obtained by stripping off a kinematic factor from the Veneziano amplitude, where the latter contains information about the polarization of external states\footnote{The polarisation information is roughly identified with the factor $I$ in \eqref{Idefin}.}. 
This is best given in terms of colour-ordered amplitudes and, in our conventions, takes the form
\begin{equation}\label{colordressed}
\begin{split}
& \mathcal{V}_{\text{open}}^{I_{1}I_{2}I_{3}I_{4}}=\frac{1}{2}\sum_{\mathcal{P}(2,3,4)}\Tr [T^{I_{1}}T^{I_{2}}T^{I_{3}}T^{I_{4}}]\mathcal{V}_{\text{open}}(1234)= \\
& \Tr [T^{I_{1}}T^{I_{2}}T^{I_{3}}T^{I_{4}}]\mathcal{V}_{\text{open}}(1234)+\Tr [T^{I_{1}}T^{I_{4}}T^{I_{2}}T^{I_{3}}]\mathcal{V}_{\text{open}}(1423)+\Tr [T^{I_{1}}T^{I_{3}}T^{I_{4}}T^{I_{2}}]\mathcal{V}_{\text{open}}(1342),
\end{split}
\end{equation}
where $\mathcal{P}(2,3,4)$ are permutations of points $(2,3,4)$, and in the second equality we exploited the antisymmetry of the $SO(N)$ generators to reduce the number of independent color traces from 6 to 3.
The colour-ordered amplitude $\mathcal{V}_{\text{open}}(1234)$ takes the form (see e.g. \cite{Schlotterer:2012ny}):
\begin{equation}
\label{venez1}
\mathcal{V}_{\text{open}}(1234)= P \exp \left( \sum_{m \geq 1} \zeta_{2m+1} M_{2m+1} \right)  \mathcal{V}_{\text{YM}}(1234), \quad \mathcal{V}_{\text{YM}}(1234)= -\frac{1}{s \, t}
\end{equation}
where
\begin{align}
& P = \exp \left( \sum_{m \geq 1}\frac{ \zeta_{2m}}{2m} \alpha'^{2m}\left( s^{2m}+t^{2m}- u^{2m} \right)  \right), \notag \\
& M_{2m+1}= \frac{1}{2m+1}  \alpha'^{2m+1}	\left( s^{2m+1}+t^{2m+1}+u^{2m+1}  \right),
\end{align}
$\mathcal{V}_{\text{YM}}(1234)$ is the colour-ordered Yang-Mills amplitude and  $s,t,u$ are 8-dimensional Mandelstam variables satisfying the on-shell constraint $s+t+u=0$.
Another equivalent way of writing the Veneziano amplitude is in terms of $\Gamma$ functions:
\begin{equation}
\mathcal{V}_{\text{open}}(1234)=\mathcal{V}_{\text{YM}}(1234) \frac{\Gamma(1-\alpha' s) \Gamma(1-\alpha' t)}{\Gamma(1+\alpha'u)}.
\end{equation}
The Veneziano amplitude satisfies many interesting properties. For example, there is a disentanglement of Riemann zeta functions with even and odd arguments, as it is most obvious from the form in equation \eqref{venez1}. This property is closely related to the well known fact that open and closed string amplitude amplitudes are related by a kernel:
\begin{equation}
\mathcal{V}_{\text{closed}}=\left( \mathcal{V}_{\text{open}}(1234) \right)^2 S
\end{equation}
where $\mathcal{V}_{\text{closed}}$ is the Virasoro-Shapiro amplitude, i.e. the amplitude of four closed strings\footnote{As for the Veneziano amplitude, with a slight abuse of language, we will refer to Virasoro-Shapiro amplitude as to the amplitude obtained by stripping off a kinematic factor from the tree-level four-point amplitude in type IIB string theory.},
and the kernel $S$ is defined by
\begin{equation}
S = \frac{1}{\pi \alpha'}\frac{\sin (\pi \alpha' s ) \sin (\pi \alpha' t )}{ \sin (\pi \alpha' u )} =  \frac{st}{u} P^{-2}.
\end{equation}
Notice that $S$ provides a cancellation of all even zetas, as it should be, since the Virasoro-Shapiro amplitude only contains odd zetas.
Relations like the one above are also known as Kawai-Lewellen-Tye (KLT) relations \cite{Kawai:1985xq}. At low-energy, they yield a relation between gluon and graviton amplitudes.

For completeness, let us also recall that colour-ordered amplitudes are related each other by further relations, for example,
\begin{equation}
\label{WSmonodromy}
\mathcal{V}_{\text{open}}(1342)= \frac{\sin (\pi \alpha' t)}{\sin (\pi \alpha' u)}\mathcal{V}_{\text{open}}(1234),
\end{equation}
which can be derived from monodromy properties of the string world-sheet \cite{Stieberger:2009hq,Bjerrum-Bohr:2009ulz}.
In the field-theory limit, i.e. $\alpha' \rightarrow 0$, they reduce to the well known Bern-Carrasco-Johansson (BCJ) relations \cite{Bern:2008qj} between colour-ordered amplitudes.

\subsection{The AdS$_5 \times$S$^3$ effective action}
The general idea is to write down an effective action starting from the Veneziano amplitude and uplift it to AdS$_5 \times$S$^3$.
Thus, let us expand the flat-space colour-ordered Veneziano amplitude, at the first few orders in $\alpha'$ 
\begin{equation}
\begin{split}
\mathcal{V}_{\text{open}}(1234)= & -\frac{1}{s\, t} + \frac{\pi^2}{6} \alpha'^2 +(s+t) \zeta_3 \alpha'^3 +\frac{\pi^4 }{720} (7 s^2 + 7 t^2 + u^2)\alpha'^4 + \\
& + \frac{1}{3}\left( \left(\frac{1}{6}\zeta_3 \pi^2+\zeta_5\right)  (s^3 + t^3)+\left(\frac{1}{6}\zeta_3 \pi^2-2\zeta_5\right) u^3  \right) \alpha'^5+ \cdots,
\end{split}
\label{eq:col_ordered_veneziano}
\end{equation}
where the even zetas have been evaluated. Excluding the field theory term the remaining expression can be viewed as arising from a scalar effective action, with contact terms containing an increasing number of derivatives. The first few terms take the form
\begin{equation}
\begin{split}
V_{\text{open}}=& \frac{1}{8}\frac{\pi^2}{6} \Tr[\phi^4]\alpha'^2 +\frac{1}{2} \zeta_3 \Tr \left[ (\partial_\mu \phi) (\partial_\mu \phi)  \phi \phi \right]\alpha'^3 + \\
&+ \frac{1}{2} \frac{\pi^4 }{720} \Tr \left[14(\partial_\mu \partial_\nu \phi) (\partial_\mu \partial_\nu \phi)\phi^2+(\partial_\mu \partial_\nu \phi)\phi (\partial_\mu \partial_\nu \phi)\phi\right]\alpha'^4+\\
& +\frac{1}{3}\Tr \left[2 \left(\frac{1}{6}\zeta_3 \pi^2+\zeta_5\right)(\partial_\mu \partial_\nu \partial_\rho \phi) (\partial_\mu \partial_\nu  \partial_\rho \phi)\phi^2+ \right.\\
& +\left. \left(\frac{1}{6}\zeta_3 \pi^2-2\zeta_5\right)(\partial_\mu \partial_\nu \partial_\rho \phi) \phi(\partial_\mu \partial_\nu  \partial_\rho \phi)\phi\right] \alpha'^5+ \cdots,
\end{split}
\end{equation} 
where the field $\phi\equiv \phi^I T^I$ is valued in the adjoint of the gauge group $SO(8)$.
It is easy to see that after going to momentum space these higher derivative contact terms provide polynomials in the Mandelstam variables and we recover the full colour-dressed amplitude   \eqref{colordressed} with the partial-ordered amplitudes as given in  \eqref{eq:col_ordered_veneziano}, order by order in $\alpha'$.

Following \cite{Abl:2020dbx}, let us now uplift this potential to an AdS$_5 \times$S$^3$ background, replacing partial flat derivatives with their AdS$_5 \times$S$^3$ covariant\footnote{Let us stress that these are AdS$_5 \times$S$^3$ covariant derivatives, they are \emph{not} covariant with respect to the gauge/global group $SO(8)$.} counterparts.
In doing so, we should note however that the uplift is not unique. This is essentially due to two reasons. 
First, the covariant derivatives no longer commute and therefore the way of arranging the derivatives in the action is ambiguous.
Secondly, at any order in $\alpha'$ there will be terms involving lower numbers of derivatives - that appeared at previous orders - compensated by the AdS (and S) radius $R$, which would vanish in the flat-space limit, i.e. the limit in which the radius of both the non-compact and the compact space is large. These ambiguities  can only be fixed by other methods.
This is a very important point which we will better clarify later on, and that is completely analogous to the AdS$_5 \times$S$^5$ case\footnote{See also \cite{Aprile:2020mus}, where similar statements were found from a CFT perspective.} \cite{Abl:2020dbx}:
the procedure can intrinsically only access the AdS$_5 \times$S$^3$ {\it completion} of the flat amplitude, while it cannot probe true curvature effects, that manifest themselves in the form of ambiguities. 
Here, with {\it completion} we mean \emph{the largest sub-amplitude which directly descends from flat space}.
As will become clear after introducing Mellin space, this notion of flat-space limit is intrinsically an extension - which takes into account the full AdS$\times$S geometry -  of the more familiar flat-space limit  \cite{Penedones:2010ue,Fitzpatrick:2011hu} and it automatically reduces to the latter when all but AdS variables are set to zero.
From now on we will use the terminology AdS$_5 \times$S$^3$ completion as in the sense explained above, i.e. the largest AdS$_5 \times$S$^3$ sub-amplitude surviving the large radius limit.

In summary, the AdS$_5 \times$S$^3$ potential has the following form\footnote{The various normalisations are chosen so that the normalisation of the associated Mellin amplitudes matches with the corresponding flat amplitude coefficient, as we will see later on.}
\begin{equation}
\begin{split}
V_{\text{AdS}_5\times \text{S}^3}^{\text{open}}=& \frac{1}{8} A \Tr[\phi^4]\alpha'^2 + \frac{1}{8} B \Tr \left[ (\nabla_\mu \phi) (\nabla_\mu \phi)  \phi \phi \right]\alpha'^3+ \\
&+\frac{1}{16} \Tr \bigl[C(\nabla_\mu \nabla_\nu \phi) (\nabla_\mu \nabla_\nu \phi)\phi^2+\frac{1}{2} D(\nabla_\mu \nabla_\nu \phi)\phi (\nabla_\mu \nabla_\nu \phi)\phi\bigr]\alpha'^4+\\
&+\Tr  \left[  E \,  \nabla^2 (\nabla_\mu \phi) (\nabla_\mu \phi)  \phi \phi+F \, \nabla^2 (\nabla_\mu \phi)\phi (\nabla_\mu \phi) \phi\right]\alpha'^4+\\
& + \frac{1}{32} \Tr \bigl[G(\nabla_\mu \nabla_\nu \nabla_\rho \phi) (\nabla_\mu \nabla_\nu  \nabla_\rho \phi)\phi^2+ \\
& +\frac{1}{32} H(\nabla_\mu \nabla_\nu \nabla_\rho \phi) \phi(\nabla_\mu \nabla_\nu  \nabla_\rho \phi)\phi\bigr] \alpha'^5+ \cdots,
\end{split}
\end{equation} 
with
\begin{align}\label{expanscoeff}
& A(\alpha')= \frac{\pi^2}{6} + a_1 \frac{\alpha'}{R^2}+ a_2 \left(\frac{\alpha'}{R^2}\right)^2+\cdots \notag \\
& B(\alpha')= \zeta_3+ b_1 \frac{\alpha'}{R^2}+ b_2 \left(\frac{\alpha'}{ R^2}\right)^2 +\cdots \notag \\
& C(\alpha')= \frac{\pi^4 }{720} 7+ c_1 \frac{\alpha'}{R^2}+ c_2 \left(\frac{\alpha'}{R^2}\right)^2+\cdots  \notag \\ 
& D(\alpha')= \frac{\pi^4 }{720}+ d_1 \frac{\alpha'}{R^2}+ d_2 \left(\frac{\alpha'}{R^2}\right)^2+\cdots  \notag \\ 
& E(\alpha')=  e_1 + e_2 \left(\frac{\alpha'}{R^2}\right)+\cdots  \notag \\ 
& F(\alpha')=  f_1 + f_2 \left(\frac{\alpha'}{R^2}\right)+\cdots \notag  \\ 
& G(\alpha')= \frac{1}{3} \left(\frac{1}{6}\zeta_3 \pi^2+\zeta_5\right)+ g_1 \frac{\alpha'}{R^2}+ g_2 \left(\frac{\alpha'}{R^2}\right)^2+\cdots \notag  \\ 
& H(\alpha')= \frac{1}{3} \left(\frac{1}{6}\zeta_3 \pi^2-2\zeta_5\right)+ h_1 \frac{\alpha'}{R^2}+ h_2 \left(\frac{\alpha'}{R^2}\right)^2+\cdots. 
\end{align}
Notice that the only terms which non trivially contribute to the AdS$_5 \times$S$^3$ completion are $A,B,C,D,G,H$ while $E,F$ are novel AdS terms that vanish upon taking the limit. We will refer to the latter as ambiguities.
From now on, we will set $R=1$.

In the next section we will compute AdS$_5 \times$S$^3$ Witten diagrams associated to this action which will provide a prediction for the four-point function of arbitrary KK modes order by order in $\alpha'$.

\subsection{AdS$\times$S Witten diagrams in embedding space}
\label{sec:Outline}
In the remaining part of the section, we will generalise the formulae of \cite{Abl:2020dbx}, which are valid for general AdS$_{\theta+1} \times$S$^{\theta+1}$ backgrounds, to the most general AdS$_{\theta_1+1} \times$S$^{\theta_2+1}$ theory. We will keep the dimensions $\theta_1, \theta_2$ generic for most of the discussion. The formulae for the AdS$_5\times$S$^3$ case can then be recovered by taking  $\theta_1=4$, $\theta_2=2$.

We will make use of the embedding space formalism.
Here, bulk points in AdS$_{\theta_1+1}$ and S$^{\theta_2+1}$ are defined via
\begin{align}
& \hat{X}^2 = - \left( \hat{X}^{-1} \right)^2 - \left( \hat{X}^{0} \right)^2+\sum_{i=1}^{\theta_1} \left(\hat{X}^i\right)^2 =-1,  \notag \\
&\hat{Y}^2 = \sum_{i=-1}^{\theta_2} \left(\hat{Y}^i\right)^{2} =1.
\end{align}
On the other hand, boundary coordinates satisfy $X^2=Y^2=0$ and are related to spacetime ($x_{ij}^2$) and $SO(4)$ distances ($y_{ij}^2$) via:
\begin{equation}
x_{ij}^2=-2 X_i \cdot X_j, \qquad \qquad   y_{ij}^2=-2 Y_i \cdot Y_j.
\end{equation}

In embedding coordinates the action of covariant derivatives can be conveniently defined in terms of projectors. These read:
\begin{equation}
\mathcal{P}_A^B = \delta_A^B + \hat{X}_A \hat{X}^B,  \qquad \mathcal{P}_I^J = \delta_I^J - \hat{Y}_I \hat{Y}^J.
\end{equation}
Note that the bulk coordinates are in the kernel of the respective projectors:
\begin{align}
& \mathcal{P}_A^B \hat{X}^A = 0, \\
& \mathcal{P}_I^J \hat{Y}^I = 0.
\end{align}
We will be particularly interested in the covariant derivative of a rank-$N$ tensor defined by \cite{Penedones:2010ue,Sleight:2016hyl}
\begin{equation}
\nabla_A T_{A_1 A_2 \ldots A_N}= \mathcal{P}_A^{C} \mathcal{P}_{A_1}^{C_1} \ldots  \mathcal{P}_{A_N}^{C_N} \partial_C \mathcal{P}_{C_1}^{E_1} \ldots  \mathcal{P}_{C_N}^{E_N} T_{E_1 \ldots E_N}.
\end{equation}

In this notation, an AdS contact Witten diagram in embedding space reads
\begin{equation}
D_{\Delta_1 \Delta_2 \Delta_3 \Delta_4}^{(\theta_1)}(X_i)= \frac{1}{(-2)^{2\Sigma_{\Delta}}} \int_{\text{AdS}} \frac{d^{\theta_1+1} \hat{X}}{P_1^{\Delta_1}P_2^{\Delta_2}P_3^{\Delta_3}P_4^{\Delta_4}}, \qquad P_i= \hat{X}\cdot X_i,
\end{equation}
where we have defined $\Sigma_{\Delta}= (\Delta_1+\Delta_2+\Delta_3+\Delta_4)/2$.

Analogously, one can define sphere contact diagrams as
\begin{equation}
B^{(\theta_2)}_{p_1 p_2 p_3 p_4}(Y_i)= (-2)^{2\Sigma_p} \int_{\text{S}} d^{\theta_2+1 }\hat{Y} Q_1^{p_1}Q_2^{p_2}Q_3^{p_3}Q_4^{p_4}\,, \qquad Q_i = \hat{Y} \cdot Y_i,
\end{equation}
where we defined $\Sigma_{p}= (p_1+p_2+p_3+p_4)/2$. Note, this last integral can be explicitly evaluated.
After some combinatorics one gets \cite{Abl:2020dbx,Chen:2020ipe}:
\begin{equation}
B^{(\theta_2)}_{p_1 p_2 p_3 p_4}(Y_i)= \mathcal{N}_S\sum_{\{d_{ij}\}} \prod_{i < j} \frac{(Y_i \cdot Y_j)^{d_{ij}}}{\Gamma[d_{ij}+1]}
\end{equation}
where the sum runs over the set 
$$
\left\{ (d_{12},d_{13},d_{14},d_{23},d_{24},d_{34}): 0 \leq d_{ij}=d_{ji}, \quad d_{ii}=0, \quad \sum_{i=1}^4 d_{ij}=p_j \right\},
$$
and the factor  $\mathcal{N}_S$ reads
\begin{equation}
\mathcal{N}_S = 2 \cdot 2^{\Sigma_p} \frac{\pi^{\theta_2 /2 +1} \prod_i \Gamma(p_i+1)}{\Gamma(\Sigma_p + \theta_2/2+1)}.
\end{equation}

As we mentioned already, the main idea will be to work directly in the product geometry AdS$\times$S. In this context it is therefore natural to define   
\begin{equation}
\mathcal{W}_{i} \equiv \frac{1}{(P_i+Q_i)},
\end{equation}
which is related to the {\it generalised bulk-to-boundary propagator} via \cite{Abl:2020dbx}
\begin{equation}
G(X_i,\hat{X};Y_i, \hat{Y})=\mathcal{C}_{\Delta_i} \frac{1}{(-2)^{\Delta_i}} \mathcal{W}_{i}^{\Delta_i},
\end{equation}
with the normalisation given by
\begin{equation}
\mathcal{C}_{\Delta}= \frac{\Gamma(\Delta)}{2\pi^{\frac{\theta_1+\theta_2}{4}}\Gamma(\Delta-\frac{\theta_1+\theta_2}{4}+1)}.
\end{equation}
Note that when $\Delta=(\theta_1+\theta_2)/2$ the generalised propagator obeys the following equation
\begin{equation}
\label{massprop}
\nabla^2 \mathcal{W}_{i}^{\Delta_i} \equiv \nabla_{\hat{X}}^2  \mathcal{W}_{i}^{\Delta_i}+ \nabla_{\hat{Y}}^2 \mathcal{W}_{i}^{\Delta_i}= \Delta_i \frac{\theta_2-\theta_1}{2}  \mathcal{W}_{i}^{\Delta_i},
\end{equation}
therefore its ``mass" is controlled by the difference between the non-compact and the compact space dimensions $\theta_1-\theta_2$. 
In particular, for $\mathcal{N}=4$ SYM it satisfies a massless equation, while in AdS$_5\times$S$^3$ it satisfies the equation for a massive scalar field.

Starting from the generalised bulk-to-boundary propagator, we then define generalised AdS$\times$S contact Witten diagrams via
\begin{equation}
D_{\Delta_1 \Delta_2 \Delta_3 \Delta_4}^{\text{AdS}_{\theta_1+1} \times \text{S}^{\theta_2+1}}(X_i,Y_i)=\frac{1}{(-2)^{2\Sigma_\Delta}} \int_{\text{AdS}_{\theta_1+1} \times \text{S}^{\theta_2+1}} d^{\theta_1+1}\hat{X} d^{\theta_2+1} \hat{Y} \,\, \mathcal{W}_{1}^{\Delta_1} \mathcal{W}_{2}^{\Delta_2} \mathcal{W}_{3}^{\Delta_3} \mathcal{W}_{4}^{\Delta_4}.
\label{genWitten}
\end{equation}

Note that AdS$\times$S Witten diagrams reduce to a sum of familiar AdS Witten diagrams after Taylor-expanding the propagators. More precisely, we have:
\begin{equation}
D_{\Delta_1 \Delta_2 \Delta_3 \Delta_4}^{\text{AdS}_{\theta_1+1} \times \text{S}^{\theta_2+1}}(X_i,Y_i)=  \sum_{p_i=0}^{\infty}\prod_{i=1}^4 (-1)^{p_i} \frac{(p_i+1)_{\Delta_i-1}}{\Gamma(\Delta_i)} D^{(\theta_1)}_{p_1 +\Delta_1 p_2+\Delta_2 p_3+\Delta_3 p_4+\Delta_4}(X_i) B^{(\theta_2)}_{p_1p_2p_3p_4}(Y_i).
\label{eq:AdS_S_Wittten}
\end{equation}

\subsection{Generalised Mellin space}
\label{sec:Mellin}
We conclude the section by defining a generalised Mellin space representation. This is a very natural generalisation of the familiar Mellin space formalism for AdS amplitudes \cite{Mack:2009mi,Penedones:2010ue} to the full AdS$\times$S product space \cite{Abl:2020dbx,Aprile:2020luw}. As we will see, the correlators admit very simple expressions when written in this formalism.

Let us first recall the Mellin transform of standard contact Witten diagrams \cite{Penedones:2010ue}:
\begin{equation}
D^{(\theta_1)}_{\Delta_1 \Delta_2 \Delta_3 \Delta_4}(X_i) = \frac{\frac{1}{2} \pi^{\theta_1/2}\Gamma(\Sigma_{\Delta}-\theta_1/2)}{(-2)^{\Sigma_\Delta}\prod_i \Gamma(\Delta_i)} \int \frac{d \delta_{ij}}{(2\pi i)^2} \prod_{i<j} \frac{\Gamma(\delta_{ij})}{(X_i \cdot X_j)^{\delta_{ij}}}, \quad \sum_i \delta_{ij}=\Delta_j.
\label{eq:D_Mellin} 
\end{equation}
As we have seen, the sphere function $B^{(\theta_2)}_{p_1p_2p_3p_4}(Y_i)$ also admits a similar representation that can be obtained by explicitly evaluating the sphere integral. Let us recall it here for convenience:
\begin{equation}
B^{(\theta_2)}_{p_1p_2p_3p_4}(Y_i) = 2 \cdot 2^{\Sigma_p} \frac{\pi^{\theta_2 /2 +1} \prod_i \Gamma(p_i+1)}{\Gamma(\Sigma_p + \theta_2/2+1)} \sum_{\{ d_{ij}\}}\prod_{i<j} \frac{(Y_i \cdot Y_j)^{d_{ij}}}{\Gamma(d_{ij}+1)}, \quad  \sum_i d_{ij}=p_j.
\end{equation}
Inserting the above representations into \eqref{eq:AdS_S_Wittten} we get
\begin{align}
&D_{\Delta_1 \Delta_2 \Delta_3 \Delta_4}^{\text{AdS}_{\theta_1+1} \times \text{S}^{\theta_2+1}}(X_i,Y_i)=\\ 
&\frac{\pi^{\frac{\theta_1+\theta_2}{2}+1}}{(-2)^{\Sigma_\Delta}\prod_i \Gamma(\Delta_i)} \sum_{p_i=0}^{\infty}(-1)^{\Sigma_p} \int \frac{d \delta_{ij}}{(2\pi i)^2}\sum_{\{ d_{ij}\}} \prod_{i<j}\left( \frac{(Y_i \cdot Y_j)^{d_{ij}}}{(X_i \cdot X_j)^{\delta_{ij}}}\frac{\Gamma(\delta_{ij})}{\Gamma(d_{ij}+1)} \right) (\Sigma_p+\theta_2/2+1)_{\Sigma_\Delta-\frac{\theta_1+\theta_2}{2}-1}.
\label{eq:Product_Witten}
\end{align}
This suggests to define the generalised Mellin transform $\mathcal{M}[f]$ of a generic function $f(X_i,Y_i)$ via
\begin{align}
&f(X_i,Y_i) \equiv \notag \\ 
& \frac{\pi^{\frac{\theta_1+\theta_2}{2}+1}}{(-2)^{\Sigma_\Delta}} \left(\prod_i\frac{\mathcal{C}_{\Delta_i}}{ \Gamma(\Delta_i)}\right) \sum_{p_i=0}^{\infty}(-1)^{\Sigma_p} \int \frac{d \delta_{ij}}{(2\pi i)^2}\sum_{\{ d_{ij}\}} \prod_{i<j}\left( \frac{(Y_i \cdot Y_j)^{d_{ij}}}{(X_i \cdot X_j)^{\delta_{ij}}}\frac{\Gamma(\delta_{ij})}{\Gamma(d_{ij}+1)} \right) \mathcal{M}[f],
\label{eq:mellin_transfrom}
\end{align}
where we have $\sum_{i\neq j}\delta_{ij}=p_j+\Delta_j$ and $\sum_{i\neq j}d_{ij}=p_j$. Using this definition, the Mellin transform of a generalised Witten diagram reads
\begin{equation}
\label{wittengen}
\mathcal{M} \left[ \left( \prod_i \mathcal{C}_{\Delta_i}\right) D_{\Delta_1 \Delta_2 \Delta_3 \Delta_4}^{AdS_{\theta_1+1} \times S^{\theta_2+1}}(X_i,Y_i)  \right] =(\Sigma_p+\theta_2/2+1)_{\Sigma_\Delta-\frac{\theta_1+\theta_2}{2}-1}.
\end{equation}

Finally, when considering higher derivative corrections it will be necessary to compute a decorated version of \eqref{genWitten} of the generic form 
\begin{equation}
\prod_i \mathcal{C}_{\Delta_i}\frac{\left( \prod_{i<j} (X_i \cdot X_j)^{n_{ij}^X}(Y_i \cdot Y_j)^{n_{ij}^Y}\right)}{(-2)^{2\Sigma_\Delta}}\int_{\text{AdS} \times \text{S}} d^{\theta_1+1}\hat{X} d^{\theta_2+1} \hat{Y}\prod_{i=1}^4 \frac{P_i^{n_i^P} Q_i^{n_i^Q}(\Delta_i)_{n_i}}{(\mathcal{W})^{\Delta_i+n_i}}.
\label{eq:dec_mel}
\end{equation}
In appendix \ref{app:Witten} we show that the generalised Mellin expression for this expression is given by
\begin{align}
\label{decoratemellin}
&\mathcal{M}[(\ref{eq:dec_mel})] = (-2)^{\Sigma_X}(2)^{\Sigma_Y} (-)^{2\Sigma_{Q}}  \left(\prod_{i=1}^4 \left( p_i+n_i^X+\Delta_i \right)_{n_i^P} ( p_i-n_i^Q-n_i^Y+1 )_{n_i^Q} \right)  \times \notag\\
&  \left( \prod_{i<j}(\delta_{ij})_{n_{ij}^X} (d_{ij}-n^Y_{ij}+1)_{n_{ij}^Y}\right)(\Sigma _{p}-\Sigma _Y+\frac{\theta _2}{2}+1)_{\Sigma _{\Delta }-\frac{\theta _1+\theta _2}{2}-1 +\Sigma _X+\Sigma _Y},
\end{align}
with $\sum_{i\neq j}\delta_{ij}=p_j+\Delta_j$ and $\sum_{i\neq j}d_{ij}=p_j$. Here, following the conventions of \cite{Abl:2020dbx}, we introduced the notation $\Sigma_Q$ for half the sum over the $n_i^Q$; $\Sigma_X$, $\Sigma_Y$ for the sum over the $n_{ij}^X$, $n_{ij}^Y$ respectively; $n_i^X =\sum_{i \neq j} n_{ij}^X$, $n_i^Y =\sum_{i \neq j} n_{ij}^Y$ and $n_i = n_i^X + n_i^Y +n_i^P + n_i^Q$.
Now that all main ingredients are in place we are ready to show some explicit results.

\section{Explicit results at orders $\alpha'^{2,3,4,5}$}
\label{sec:expliv}
Having presented the definitions for arbitrary $(\theta_1,\theta_2)$ we now specialise to the case at hand, tree-level $\alpha'$ correction in AdS$_5 \times$S$^3$, where we set $\theta_1=4$, $\theta_2=2$, $\Delta_i=3$. We begin by considering the constraints imposed on the $(\delta_{ij},d_{ij})$ i.e.
\begin{align}\label{constr1}
 \sum_{i\neq j}\delta_{ij}=p_j+1; \quad \quad \sum_{i\neq j}d_{ij}=p_j-2,
\end{align}
where here we have employed the shift $p_i \rightarrow p_i-2$ due to the lowest correlator being $p_i=2$. While the correlator may be written in terms of the constrained $(\delta_{ij},d_{ij})$, we will find it useful to partially solve these constraints. Following analogous conventions to those in \cite{Aprile:2022tzr}, we can write
\begin{align}
& \delta_{12}= -s+p_{(12)}, \qquad \delta_{34}= -s+p_{(34)}, \notag \\
& \delta_{14}=  -t+p_{(14)}, \qquad \delta_{23}= -t+p_{(23)}, \notag  \\
& \delta_{13}= -u+p_{(13)}, \qquad \delta_{24}= -u+p_{(24)}, \notag \\
& d_{12}= \tilde{s}+p_{(12)}, \qquad  d_{34}= \tilde{s}+p_{(34)}, \notag \\
& d_{14}= \tilde{t}+p_{(14)}, \qquad  d_{23}= \tilde{t}+p_{(23)}, \notag \\
& d_{13}= \tilde{u}+p_{(13)}, \qquad  d_{24}= \tilde{u}+p_{(24)},
\end{align}
where we introduced the notation $p_{(ij)}=\frac{ p_i+p_j}{2}$. Having done so, we are left with only 6 variables which satisfy
\begin{align}
 s+t+u= \Sigma_p -1; \quad \quad \tilde{s}+\tilde{t}+\tilde{u}= -\Sigma_p -2.
\end{align}
It follows that, at given order in $\alpha'$, the correlator will depend on 8 \emph{unconstrained} variables: $s,t,\tilde{s},\tilde{t}$ as well as the charges $p_i$. 

Now, the large $p$ limit \cite{Aprile:2020luw} suggests to trade the Mellin variables $s,t,u$ with the bold-face variables defined via:
\begin{equation}
{\bf s}=s+\tilde{s}, \qquad {\bf t}=t+\tilde{t},
\qquad {\bf s}+{\bf t}+{\bf u}=-3.
\end{equation}
In fact, as we will see, the various correlators will admit a natural stratification w.r.t to the bold-face variables ${\bf s},{\bf t}$. In particular, as  explained in \cite{Aprile:2020luw}, in the limit of large $s,t,\tilde{s},\tilde{t},p_i$ variables, the correlator approaches to the flat S-matrix with the Mandelstam replaced by the bold-face variables.
In conclusion, we will express the correlators in terms of the set (${\bf s},{\bf t},\tilde{s},\tilde{t},p_1,p_2,p_3,p_4$).

Finally, our notation for the $\alpha'$ expansion of the Mellin amplitudes will closely follow the analogous position space expression \eqref{positionspaceexp}, i.e.\footnote{From now on, we will suppress the subscript $\vec{p}$ for the Mellin amplitudes as we will always be referring to the individual correlators.}
\begin{equation}
\mathcal{M}_{\text{tree}}^{I_1 I_2 I_3 I_4}=\mathcal{M}_{\text{YM}}^{I_1 I_2 I_3 I_4}+ \mathcal{M}_{0}^{I_1 I_2 I_3 I_4}\alpha'^2+\mathcal{M}_{1}^{I_1 I_2 I_3 I_4}\alpha'^3+\mathcal{M}_{2}^{I_1 I_2 I_3 I_4}\alpha'^4+\mathcal{M}_{3}^{I_1 I_2 I_3 I_4}\alpha'^5+\cdots,
\end{equation}
where the full colour-dressed amplitude $\mathcal{M}_{\text{tree}}^{I_1 I_2 I_3 I_4}$ admits a decomposition in terms of colour-ordered amplitudes
\begin{equation}\label{colourMell}
\begin{split}
& \mathcal{M}_{\text{tree}}^{I_1 I_2 I_3 I_4}=\frac{1}{2}\sum_{\mathcal{P}(2,3,4)}\Tr [T^{I_{1}}T^{I_{2}}T^{I_{3}}T^{I_{4}}]\mathcal{M}_{\text{tree}}(1234)= \\
& \Tr [T^{I_{1}}T^{I_{2}}T^{I_{3}}T^{I_{4}}]\mathcal{M}_{\text{tree}}(1234)+\Tr [T^{I_{1}}T^{I_{4}}T^{I_{2}}T^{I_{3}}]\mathcal{M}_{\text{tree}}(1423)+\Tr [T^{I_{1}}T^{I_{3}}T^{I_{4}}T^{I_{2}}]\mathcal{M}_{\text{tree}}(1342),
\end{split}
\end{equation}

\subsection{Field-theory correlator ($\alpha'=0$)}

Before presenting our new results, for completeness let us recall the form of the correlator in the field theory limit, first computed in \cite{Alday:2021odx}, within this formalism.
This cannot be directly recovered from the effective action as it is not a polynomial in the Mandelstam and it would require adding a non-local vertex to the action.

However, this is a very special case. In fact, the correlators at this order obey a hidden $8d$ conformal symmetry \cite{Alday:2021odx}, which implies that the correlator for arbitrary KK modes can be obtained from a generating function which only depends on $8d$ distances. This generating function takes the form of the field-theory correlator with lowest charges:
\begin{equation}
\label{pqrsfieldtheory}
\sum_{\vec{p}} G_{\text{YM},\vec{p}} (x_{ij}^2+y_{ij}^2) \propto \frac{1}{(x_{12}^2+y_{12}^2)^3 (x_{34}^2+y_{34}^2)^3} D_{2321}(x_{ij}^2+y_{ij}^2).
\end{equation}
Note that this object transforms as the four-point function of weight 3  (i.e. the dimension of scalar in $8d$) operators.

Let us see how to recover this from the point of view of generalised Witten diagrams.
Note that the Pochhammer in \eqref{wittengen} vanishes when
\begin{equation}
\label{adsxs=s}
\Sigma_\Delta = \frac{\theta_1+\theta_2}{2}+1.
\end{equation}
It is immediate to see that, when \eqref{adsxs=s} is satisfied - which is the case of the field theory correlator\footnote{In fact, for the field-theory correlator we have $1/2(2+3+2+1)=(4+2)/2+1$.} - an AdS$\times$S contact diagram \eqref{genWitten} becomes proportional to a standard AdS Witten diagram \eqref{eq:D_Mellin} with the replacement $X_i\cdot X_j\rightarrow X_i\cdot X_j+Y_i\cdot Y_j$. This is nothing but \eqref{pqrsfieldtheory} written in embedding coordinates.

The generalised Mellin space expression is also very simple and can be written in the form \eqref{eq:mellin_transfrom} with $\Delta=3$ and the Mellin amplitude:
\begin{equation}
\label{fieldtheoryamp}
\mathcal{M}_{\text{YM}} (1234)=- \frac{1}{(\delta_{34}-d_{34}-1)(\delta_{14}-d_{14}-1)}=- \frac{1}{({\bf s} +1)\,( {\bf t}+1)}
\end{equation}
which is in agreement with the expression given in \cite{Alday:2021odx}. In the second line we have expressed the amplitude in terms of the bold-face variables ${\bf s}, {\bf t}$ defined above, as in \cite{Drummond:2022dxd}.
Nicely, these colour-ordered amplitudes satisfy BCJ relations for \emph{all} Kaluza-Klein operators completely analogously to flat space \cite{Drummond:2022dxd}.
In fact, we have
\begin{equation}
\label{bcjads}
\mathcal{M}_{\text{YM}} (1234)= \frac{({\bf u} +1)}{({\bf t} +1)}\mathcal{M}_{\text{YM}} (1342)
\end{equation}
and similarly for other colour-ordered amplitudes.

\subsection{Order $\alpha'^2$}
The first string corrections are at order $\alpha'^2$.
The relevant term in the effective action is
\begin{equation}
S_0 = \frac{1}{8} \zeta_2 \int_{\text{AdS}_5 \times\text{S}^3} d^5 \hat{X}d^3 \hat{Y} \Tr \left[ \phi (\hat{X}, \hat{Y})^4 \right].
\end{equation}
To obtain the correlators we just mimic the standard AdS/CFT procedure, i.e. we take derivatives w.r.t the boundary data for the bulk field which acts as a source for the scalar operator $\mathcal{O}$. 
As we stressed already, the actual difference with the common AdS/CFT prescription is just the replacement of AdS bulk-to-boundary propagators with generalised ones. Thus, at this order we get 
\begin{equation}
\begin{split}
\langle \mathcal{O}\mathcal{O}\mathcal{O}\mathcal{O} \rangle |_{\alpha'^2} (1234) &= \zeta_2 \frac{\mathcal{C}_3^4}{(-2)^{12}} \int_{\text{AdS}_5 \times\text{S}^3} \frac{d^{5}\hat{X} d^{3} \hat{Y}}{\mathcal{W}_{1}^{3}\mathcal{W}_{2}^{3}\mathcal{W}_{3}^{3}\mathcal{W}_{4}^{3}}\\
&= \zeta_2 \, \mathcal{C}_3^4\, D_{3333}^{\text{AdS}_{5}\times \text{S}^{3}}(X_i,Y_i),
\end{split}
\end{equation}
and analogously for the other colour-ordered correlators.
Thus, at this order the correlator is just a single $D_{3333}^{\text{AdS}_{5}\times \text{S}^{3}}$ function. The position space expression in terms of standard AdS Witten diagrams (D-functions) for the individual correlators can be directly read off from \eqref{eq:AdS_S_Wittten}.
In particular, note that the correlator $\langle \mathcal{O}_2 \mathcal{O}_2\mathcal{O}_2\mathcal{O}_2 \rangle$ is proportional to $\bar{D}_{3333}$. As expected, this is the same function showing up as ambiguity in the corresponding field theory one-loop amplitude \cite{Huang:2023oxf}, and can be seen as a one-loop counterterm from the field-theory viewpoint.

We are interested in the Mellin space expression\footnote{Let us stress again that this is the defined for the individual correlators.} which we can get from \eqref{wittengen}. This is very simple and reads:
\begin{equation}
\label{alpha2}
\mathcal{M}_{0} (1234)= \zeta_2 \left( \Sigma_p -2 \right)_2.
\end{equation}

At this point it is worth taking a break and noticing that the correlators so computed correctly reproduce the flat-space limit  as given in \cite{Penedones:2010ue,Fitzpatrick:2011hu}, which in the following will be referred to as \emph{AdS-type limit}. 
In fact, Penedones formula states that the Mellin amplitude in the limit of large Mellin variables approaches the flat amplitude as a function of Mandelstam variables:
\begin{equation}
\label{penedoneslim}
\mathcal{M}(\delta_{ij})  \, \, \xrightarrow[\text{large} \, \delta_{ij}]{} \, \, \frac{1}{\Gamma (\Sigma_p-2)} \,\,\int_0^\infty d\alpha \, \alpha^{\Sigma_p-1}\, \, \mathcal{V}_{\text{open}} (\alpha \delta_{ij}),
\end{equation}
where $\mathcal{V}_{\text{open}}$ stands for any of the colour-ordered flat amplitudes. 
At this order in $\alpha'$ we have $\mathcal{V}=\zeta_2$ is a constant, therefore \eqref{penedoneslim} returns the full amplitude and we recover \eqref{alpha2} straight away\footnote{Note that for $\alpha'=0$ we have $\mathcal{V}_{\text{YM}}=- \frac{1}{s \, t}$ and we correctly reproduce \eqref{fieldtheoryamp} in the AdS-type limit.}.

This fact - which represents a \emph{consistency check} for our results - will be true at all orders as can be seen directly from the Mellin transform of a generic decorated integral \eqref{decoratemellin}.
In fact, note that the decorated integral at a given order produces polynomials in $X_i \cdot X_j$ whose degree is dictated by the number of AdS derivatives hitting the vertex, or, in other words, from the order in $\alpha'$.
Schematically, a vertex of the generic form $\nabla_{\mu_1}\cdots \nabla_{\mu_n} \phi \nabla_{\mu_1}\cdots \nabla_{\mu_n} \phi \phi \phi$ goes like
\begin{equation}
\nabla_{\mu_1}\cdots \nabla_{\mu_n} \phi \nabla_{\mu_1}\cdots \nabla_{\mu_n} \phi \phi \phi \qquad  \underset{\text{large} \, X}{\sim} \qquad (X_i \cdot X_j)^n \qquad \underset{\text{Mellin}}{\sim} \qquad \delta_{ij}^{n}.
\end{equation}

Thus, for large Mellin ($\delta_{ij}$) variables, \eqref{decoratemellin} approaches\footnote{We remind again that we have shifted the charges $p_i$ by two units, $p_i\rightarrow p_i-2$, and taken $\Delta_i=3$.}
\begin{equation}
\mathcal{M}  \, \, \xrightarrow[\text{large}\, \delta_{ij}]{} \, \, \delta_{ij}^{n_{ij}^X} (\Sigma_p-2)_{n_{ij}^X+2}
\end{equation}
which is precisely the Pochhammer appearing in \eqref{penedoneslim} after integrating \eqref{penedoneslim} with $\mathcal{V} \sim \delta_{ij}^{n_{ij}^X}$. At the next orders it will become clear that the AdS-type limit turns out to be a \emph{particular} case of a more general notion of flat-space limit where the {\it sphere} variables are set to zero.

\subsection{Order $\alpha'^3$}
The computation of higher order corrections is conceptually similar, but more computationally involved because it requires evaluating the action of the covariant derivatives on fields. 
We have used a straightforward generalisation of the algorithm outlined in \cite{Abl:2020dbx}. 
At $\alpha'^3$, the relevant term in the action is:
\begin{equation}
\label{actionalpha3}
S_1^{\text{main}} = \frac{1}{8} \zeta_3 \int_{\text{AdS}_5 \times\text{S}^3} d^5 \hat{X}d^3 \hat{Y} \Tr \left[ (\nabla_\mu \phi) (\nabla_\mu \phi) \phi \phi \right].
\end{equation}
Now, it is easy to see that:
\begin{equation}
(\nabla_\mu \mathcal{W}_1^\Delta) (\nabla_\mu \mathcal{W}_2^\Delta)= \Delta^2 \frac{N_{12}}{\mathcal{W}_1^{\Delta+1} \mathcal{W}_2^{\Delta+1} }
\end{equation}
where we have defined
\begin{equation}\label{nij}
N_{ij}=X_i \cdot X_j+Y_i \cdot Y_j+P_i P_j -Q_i Q_j.
\end{equation}
Using the above relation we get, for the colour-ordered correlators,
\begin{equation}
\begin{split}
\langle \mathcal{O}\mathcal{O}\mathcal{O}\mathcal{O} \rangle |^{\text{main}}_{\alpha'^3} (1234) &  = \frac{1}{8} \zeta_3 \frac{3^2 \,\mathcal{C}_3^4}{(-2)^{12}} \int_{\text{AdS}_5 \times\text{S}^3} d^{5}\hat{X} d^{3} \hat{Y}  \frac{1}{(\mathcal{W}_{1})^{3}(\mathcal{W}_{2})^{3}(\mathcal{W}_{2})^{3}(\mathcal{W}_{4})^{3}}\\
& \times 2 \left(  \frac{ N_{12}}{\mathcal{W}_{1}\mathcal{W}_{2}}+ \frac{ N_{34}}{\mathcal{W}_{3}\mathcal{W}_{4}}+ \frac{ N_{14}}{\mathcal{W}_{1}\mathcal{W}_{4}}+ \frac{ N_{23}}{\mathcal{W}_{2}\mathcal{W}_{3}}\right).
 \end{split}
\end{equation}
This is the generator of all correlators. From \eqref{eq:dec_mel}, it is straightforward to get the associated Mellin amplitude for the individual correlators. This is given by
\begin{equation}
\mathcal{M}_{1}^{\text{main}} (1234)= \zeta_3(\mathcal{M}_1^s+\mathcal{M}_1^t-3 (\Sigma_p-2)_2 ) 
\end{equation}
where we defined the $s$-type Mellin amplitude
\begin{align}\label{alpha3}
& \mathcal{M}_1^s=(\Sigma_p-2)_3 {\bf s}-3(\Sigma_p-2)_2 \tilde{s},
\end{align}
with $\mathcal{M}_1^t$, $\mathcal{M}_1^u$ related to $\mathcal{M}_1^s$ by crossing:
\begin{align}
& \mathcal{M}_1^t \equiv \mathcal{M}_1^s [ s\rightarrow t, \, \tilde{s}\rightarrow \tilde{t}, \, p_2 \leftrightarrow p_4 ]=(\Sigma_p-2)_3 {\bf t}-3(\Sigma_p-2)_2 \tilde{t}, \notag \\
& \mathcal{M}_1^u \equiv \mathcal{M}_1^s [s\rightarrow u, \, \tilde{s}\rightarrow \tilde{u}, \, p_2 \leftrightarrow p_3 ]=(\Sigma_p-2)_3 {\bf u}-3(\Sigma_p-2)_2 \tilde{u}.
\label{crossingamp}
\end{align}
Then, the full colour-ordered amplitude is the sum of the above term (the ``main" amplitude) and the ambiguities present at this order. In sum, we have\footnote{Note that we simplified the result by absorbing the constant term $3 (\Sigma_p-2)_2$ into the ambiguity $a_1$. We will use the freedom to perform such redefinitions at higher orders as well.}:
\begin{align}
\begin{split}
\mathcal{M}_{1} (1234) & = \zeta_3(\mathcal{M}_1^s+\mathcal{M}_1^t) + a_1 (\Sigma_p-2)_2  \\
& = \zeta_3\bigl( (\Sigma_p-2)_3 ({\bf s}+{\bf t})-3(\Sigma_p-2)_2( \tilde{s}+\tilde{t}) \bigr) + a_1 (\Sigma_p-2)_2,\
\end{split}
\end{align}
and similarly for the other colour-ordered amplitudes. Here $a_1$ is a free coefficient corresponding to the freedom of adding the $\alpha'^2$ correction to the amplitude, see \eqref{expanscoeff}.
This is the only ambiguity present at this order. The other possibility would be a term of the form $\sim (\nabla^2 \phi) \phi \phi \phi$ but, because of \eqref{massprop}, it is essentially the same as the ambiguity coming from $\phi^4$.

Note that we have the relation
\begin{equation}\label{adsonshell}
{\cal M}_1^s+{\cal M}_1^t+{\cal M}_1^u \propto (\Sigma-2)_2
\end{equation}
which can be identified as the AdS analogue of the flat on-shell relation $s+t+u=0$.

Before computing the other $\alpha'$ corrections, let us notice a remarkable simplification occurring for the first two corrections which was already spotted in the AdS$_5 \times$S$^5$ case \cite{Aprile:2020mus}. This will help to write the other $\alpha'$ corrections in a more compact form. The idea is to absorb the various Pochhammers appearing through a double integral transform. 
Let us thus define the following pre-amplitude
\be\label{integral_tr_James}
\mathcal{M}_n = \frac{i}{2 \pi} \int_0^\infty  d\alpha \int_\mathcal{C} d\beta\ e^{-\alpha-\beta} \alpha^{\Sigma_p-1}  (-\beta)^{2-\Sigma_p} \, \tilde{\mathcal{M}}_n(S,T,\tilde{S},\tilde{T}),
\ee
where $\mathcal{C}$ is the Hankel contour. Here $\mathcal{M}_n$ is any of the colour-ordered amplitudes and $\tilde{\cal M}_n$ is a simplified amplitude, defined in terms of the following variables,
\be \label{capSdef}
S = \alpha s - \beta \tilde{s},\quad\tilde{S} = \alpha s + \beta \tilde{s},
\ee
and similarly for $t$-type and $u$-type variables. The integral transform \eqref{integral_tr_James} really just provides the $\Gamma$ functions needed to reconstruct the various Pochhammer. In fact, it is not difficult to check that in the pre-amplitude $\tilde{\cal M}_n$ all Pochhammer disappear. For example, the pre-amplitude associated to $\alpha'^2$ is just
\begin{equation}
\tilde{\cal M}_0(1234)=\zeta_2.
\end{equation}
On the other hand, the pre-amplitude at order $\alpha'^3$ reads
\eqref{alpha3} is
\begin{align}
& \tilde{\cal M}_1 (1234)=\zeta_3(S+T+a_1).
\end{align}
Thus, from these first two examples, we can see that the pre-amplitude is given by the corresponding term in the flat Veneziano amplitude with the Mandelstam variables replaced by $S,T$ plus lower order terms in $S,T,\tilde{S},\tilde{T}$. This will be true at higher orders as well.

Note, the $\alpha$ integral is nothing but the Penedones integral \eqref{penedoneslim}, and the above transform reduces to the latter when setting all but the Mellin variables to zero.
As mentioned previously, these \emph{non-trivial simplifications} strongly suggest that the usual flat-space limit arises as a particular case of a more general flat-space limit which involves all 8 variables which the Mellin amplitude depends on.

\subsection{Order $\alpha'^4$}
Let us now consider $\alpha'^4$ corrections. The corresponding term in the action is:
\begin{equation}
S_2^{\text{main}} =\frac{1}{16} \frac{\pi^4 }{720} \int_{\text{AdS}_5 \times\text{S}^3} d^5 \hat{X}d^3 \hat{Y} \Tr \left[7(\nabla_\mu \nabla_\nu \phi) (\nabla_\mu \nabla_\nu \phi)\phi^2+\frac{1}{2}(\nabla_\mu \nabla_\nu \phi)\phi (\nabla_\mu \nabla_\nu \phi)\phi\right].
\end{equation}
Let us stress again that this is the only term at this order with a flat-space counterpart.
After computing the derivatives we get
\begin{equation}
\begin{split}
 & \langle \mathcal{O}\mathcal{O}\mathcal{O}\mathcal{O} \rangle^{\text{main}}|_{\alpha'^4} (1234)   = \frac{1}{16} \frac{\pi^4 }{720} \frac{3^2 \,\mathcal{C}_3^4}{(-2)^{12}} \int_{\text{AdS}_5 \times\text{S}^3} d^5 \hat{X} d^3 \hat{Y} \frac{1}{(\mathcal{W}_1)^3(\mathcal{W}_2)^3(\mathcal{W}_3)^3(\mathcal{W}_4)^3} \times \\ 
 &2 \left[ 7\left( \frac{L_{12}}{(\mathcal{W}_1)^2(\mathcal{W}_2)^2}+ \frac{L_{34}}{(\mathcal{W}_3)^2(\mathcal{W}_4)^2}+ \frac{L_{14}}{(\mathcal{W}_1)^2(\mathcal{W}_4)^2}+ \frac{L_{23}}{(\mathcal{W}_2)^2(\mathcal{W}_3)^2} \right)+ \frac{L_{13}}{(\mathcal{W}_1)^2(\mathcal{W}_3)^2}+\frac{L_{24}}{(\mathcal{W}_2)^2(\mathcal{W}_4)^2} \right],
 \end{split}
\end{equation}
where here we have defined 
\begin{equation}
L_{ij} = 16 N_{ij}^2 - (P_iP_j-Q_iQ_j)(3 P_i P_j -5 Q_i Q_j -P_i Q_j -P_j Q_i).
\end{equation}
From \eqref{decoratemellin}, we get the associated colour-ordered Mellin amplitude for the individual correlators:
\begin{equation}
{\cal M}_2^{\text{main}}(1234)=\frac{\pi^4 }{720}(7{\cal M}_2^s+7{\cal M}_2^t+{\cal M}_2^u),
\end{equation}
where the $s$-type amplitude reads
\begin{equation}
{\cal M}_2^s=\left(\Sigma_p-2 \right)_4 {\bf s}^2- \left(\Sigma_p-2 \right)_3 {\bf s}(8\tilde{s}+ \Sigma_p +1)+ \left(\Sigma_p-2 \right)_2 \left(12\tilde{s}^2-\frac{3}{8} P +12\tilde{s}+\frac{3}{2}\Sigma_p \right).
\end{equation}
and analogously for ${\cal M}_2^t,{\cal M}_2^u$.
Here we have defined
\begin{equation}
P \equiv p_1^2+p_2^2+p_3^2+p_4^2.
\end{equation}
As mentioned before, the amplitude automatically stratifies w.r.t the power of $\bf{s},\,\bf{t}$. In particular it respects the large $p$ behaviour \cite{Aprile:2020luw}.

By using (the inverse of) \eqref{integral_tr_James}, we can get the pre-amplitude associated to this expression. This is very simple and reads
\begin{equation}
\tilde{\cal M}_2^s= S^2 + S \Sigma_p - \frac{5}{2} S+ \frac{3}{2}\left( \tilde{S}+\Sigma_p \right) -\frac{3}{8}P.  
\end{equation}
Let us now compute the ambiguities. Through explicit calculation we verified that, on the top of the lower-order ambiguities, there are two new independent ambiguities at this order, which we choose to be
\begin{equation}
\label{amba4}
\Tr \left[( \nabla^2 \nabla_\mu \phi) (\nabla^\mu \phi) \phi \phi \right], \qquad \Tr \left[( \nabla^2 \nabla_\mu \phi)\phi (\nabla^\mu \phi)  \phi \right].
\end{equation}
After computing the derivatives and using the Mellin space formula \eqref{decoratemellin} we obtain, respectively:
\begin{equation}
\mathcal{M}_{2,\text{amb}_1}={\cal M}_{2,\text{amb}}^s+{\cal M}_{2,\text{amb}}^t,
\end{equation}
and
\begin{equation}
\mathcal{M}_{2,\text{amb}_2}={\cal M}_{2,\text{amb}}^u,
\end{equation}
with
\begin{equation}
{\cal M}_{2,\text{amb}}^s=(\Sigma_p-2)_3 {\bf s}+\frac{3}{14}  (\Sigma_p-2)_2 (p_1 p_2 +p_3 p_4+ 2\Sigma_p-2\Sigma_p^2-4\tilde{s}\Sigma_p - 2\tilde{s}),
\end{equation}
and ${\cal M}_{2,\text{amb}}^{t,u}$ related to ${\cal M}_{2,\text{amb}}^s$ by crossing, \textit{cf.} \eqref{crossingamp}.
In terms of the associated pre-amplitude we have
\begin{equation}
\tilde{\cal M}_{2,\text{amb}}^s = S+ \frac{3}{8}\left(2 \tilde{S}+ 2 \Sigma_p-2\Sigma_p^2  + p_1 p_2+ p_3 p_4 \right).
\end{equation}

In sum, the full colour-ordered (pre-)amplitude at order $\alpha'^4$ is\footnote{Here we have used a slight abuse of notation, as the ambiguities here are related to those in \eqref{expanscoeff} by some shifts and rescaling.}
\begin{equation}
\begin{split}
\tilde{\cal M}_2 (1234)= &  \frac{\pi^4 }{720}(7\tilde{\cal M}_2^s+7\tilde{\cal M}_2^t+\tilde{\cal M}_2^u)+ a_{2} + b_{1} \left(\tilde{\cal M}_1^s+\tilde{\cal M}_1^t\right)+ \\
& + e_{1} (\tilde{\cal M}_{2,\text{amb}}^s+\tilde{\cal M}_{2,\text{amb}}^t) +f_{1} \tilde{\cal M}_{2,\text{amb}}^u.
\end{split}
\end{equation}
Note that this theory will generically contain more ambiguities than $\mathcal{N}=4$ SYM \cite{Abl:2020dbx,Aprile:2020mus}, due to the loss of some crossing symmetry.

Moreover, as anticipated, it is easy to check that for large $S,T,U$ the amplitude is given by the corresponding term in the expansion of the Veneziano amplitude with the Mandelstam variables replaced by $S,T,U$ variables:
\begin{equation}
\tilde{\cal M}_2(1234) \underset{\text{large}\, S,T,U}{\sim}  \frac{\pi^4 }{720}(7 S^2+7 T^2 +U^2)= \mathcal{V}_{\text{open}}\big|_{\alpha'^4}(1234).
\end{equation}

\subsection{Order $\alpha'^5$}
Finally, let us compute the amplitude at order $\alpha'^5$. The main amplitude at this order reads
\begin{equation}
\begin{split}
S_3^{\text{main}} = & \frac{1}{3} \frac{1}{32} \int_{\text{AdS}_5 \times\text{S}^3} d^5 \hat{X}d^3 \hat{Y} \Tr \left[\left(\frac{1}{6}\zeta_3 \pi^2+\zeta_5\right)(\nabla_\mu \nabla_\nu \nabla_\rho \phi) (\nabla_\mu \nabla_\nu  \nabla_\rho \phi)\phi^2+ \right.\\
& +\left. \frac{1}{2}\left(\frac{1}{6}\zeta_3 \pi^2-2\zeta_5\right)(\nabla_\mu \nabla_\nu \nabla_\rho \phi) \phi(\nabla_\mu \nabla_\nu  \nabla_\rho \phi)\phi\right].
\end{split}
\end{equation}
In appendix \ref{app:details} we explicitly evaluate the derivatives, along with their Mellin space expression. Instead here we just give the associated Mellin pre-amplitude, which reads
\begin{equation}
\tilde{\cal M}_3^{\text{main}}=\frac{1}{3} \left(\frac{1}{6}\zeta_3 \pi^2+\zeta_5\right) (\tilde{\cal M}_3^s+\tilde{\cal M}_3^t) + \frac{1}{3}\left(\frac{1}{6}\zeta_3 \pi^2-2\zeta_5\right) 
 \tilde{\cal M}_3^u 
\end{equation}
with\footnote{We remind that $P \equiv p_1^2+p_2^2+p_3^2+p_4^2$.}
\begin{equation}
\label{alpha5main}
\begin{split}
\tilde{\cal M}_3^s = \, &  S^3 -3 S^2 \Sigma_p +2 S \Sigma_p ^2 - \frac{11}{8} P S-3 S^2 + 7 S\Sigma_p+\frac{9}{2}S\tilde{S}-3\Sigma_p \tilde{S} +\frac{33}{16}P-2\Sigma_p^2\\ +& p_1p_2+p_3p_4+ \frac{45}{2}S-\frac{25}{4}\Sigma_p -\frac{19}{4}\tilde{S},
\end{split}
\end{equation}
and $\tilde{\cal M}_3^t,\, \tilde{\cal M}_3^u$ defined analogously.

By writing down all possible ambiguities, we find that there are in total 8 independent, and we can choose them to be the 6 associated to previous orders plus two new ones associated to the terms
\begin{align}
& \text{Tr}\left[(\nabla^2 \nabla^\mu \nabla^\nu \phi) (\nabla_\nu \nabla_\mu \phi) \phi \phi\right],\qquad \text{Tr}\left[(\nabla^2 \nabla^\mu \nabla^\nu \phi)\phi (\nabla_\nu \nabla_\mu \phi) \phi \right]
\end{align}
The pre-amplitude expressions are $\tilde{\cal M}_{\text{3,amb}}^s +\tilde{\cal M}_{\text{3,amb}}^t$ and $\tilde{\cal M}_{\text{3,amb}}^u$ respectively, with the $s$-channel ambiguity given by
\begin{equation}
\label{alpha5amb}
\begin{split}
\tilde{\cal M}_{\text{3,amb}}^s =\,& S^2 - \frac{8}{5}S \Sigma^2+\frac{8}{5}\Sigma^3-\frac{2}{5}P\Sigma +\frac{2}{5}(p_1 p_2+p_3 p_4) (2S-2\Sigma+1)+\frac{8}{5}S \tilde{S}\\
&- \frac{8}{5}\tilde{S}\Sigma +\frac{3}{5}S \Sigma +\frac{39}{10}S -\frac{1}{10}\tilde{S}+\frac{3}{8}P -\frac{4}{5}\Sigma^2 -\frac{7}{10}\Sigma,
\end{split}
\end{equation}
and $\tilde{\cal M}_{\text{3}}^{t,u}$,$\tilde{\cal M}_{\text{3,amb}}^{t,u}$ defined accordingly.

In sum, the full Mellin pre-amplitude reads
\begin{equation}
\begin{split}
\tilde{\cal M}_3(1234) = &  \frac{1}{3} \left(\frac{1}{6}\zeta_3 \pi^2+\zeta_5\right) (\tilde{\cal M}_3^s+\tilde{\cal M}_3^t) + \frac{1}{3}\left(\frac{1}{6}\zeta_3 \pi^2-2\zeta_5\right) 
 \tilde{\cal M}_3^u + \\
 &+ a_3 + b_2(\tilde{\cal M}_1^s+\tilde{\cal M}_1^t)+ c_1(\tilde{ \cal M}_2^s+ \tilde{\cal M}_2^t)+d_1 \tilde{\cal M}_2^u+ e_2 (\tilde{\cal M}_{2,\text{amb}}^s+\tilde{\cal M}_{2,\text{amb}}^t) +\\
& + f_2 \tilde{\cal M}_{2,\text{amb}}^u + l_1(\tilde{\cal M}_{\text{3,amb}}^s+\tilde{\cal M}_{\text{3,amb}}^t)+ h_1\tilde{\cal M}_{3,\text{amb}}^u.
\end{split}
\end{equation}

Finally, note once again that, for large $S,T,U$, the amplitude reduces to the flat-Veneziano amplitude:
\begin{equation}
\tilde{\cal M}_3(1234) \underset{\text{large}\, S,T,U}{\sim}   \frac{1}{3} \left(\frac{1}{6}\zeta_3 \pi^2+\zeta_5\right) (S^3+T^3) + \frac{1}{3}\left(\frac{1}{6}\zeta_3 \pi^2-2\zeta_5\right) U^3 = \mathcal{V}_{\text{open}}\big|_{\alpha'^5}(1234).
\end{equation}

\subsection{Towards a generalised flat-space limit}
We conclude this section by commenting on the various flat-space limit we encountered along the way.
We would like to address the following question:

\begin{mdframed}
\center
\emph{what is the largest sub-amplitude which directly descends from flat-space}? 
\end{mdframed}

A first answer comes from the flat-space limit worked out in \cite{Penedones:2010ue,Fitzpatrick:2011hu}, which states that the Mellin amplitude and the flat scattering amplitude are related by an integral transform which, in this theory, reads
\begin{equation}
\label{penedoneslim2}
\mathcal{M}(s,t)  \, \, \xrightarrow[\text{large}\,s,t]{} \, \, \frac{1}{\Gamma (\Sigma_p-2)} \,\,\int_0^\infty  d\alpha \,e^{-\alpha}  \alpha^{\Sigma_p-1}\, \, \mathcal{V}  (\alpha s,\alpha t).
\end{equation}
Note that only AdS variables participate in this limit; the sphere variables $\tilde{s},\tilde{t}$ and the charges $p_i$ are just spectators.
This does not look completely satisfactory because in this particular theory, AdS and $S$ factors scale in the same way for large radius, thus one would expect a limit where variables are treated in a more symmetric way.

In fact, the authors of \cite{Aprile:2020luw} point out that in the ``large $p$ limit", i.e. in the limit of large $s,t,\tilde{s},\tilde{t},p_i$, the relation above gets upgraded to a more symmetric version with the Mellin amplitude related to the corresponding flat-space scattering process via the integral
\begin{equation}
\mathcal{M}( {\bf s}, {\bf t})  \, \, \xrightarrow[\text{large}\,s,t,\tilde{s},\tilde{t},p_i]{} \, \, \frac{1}{\Gamma (\Sigma_p-2)} \,\,\int_0^\infty  d\alpha \, e^{-\alpha} \alpha^{\Sigma_p-1}\, \, \mathcal{V}  (\alpha {\bf s},\alpha {\bf t}),
\end{equation}
where we recall here for convenience the definition of the bold-face variables
\begin{equation}
{\bf s}=s+\tilde{s}, \qquad {\bf t}=t+\tilde{t}, \qquad  {\bf s}+ {\bf t}+ {\bf u}=-3.
\end{equation} 
Note that all Mellin amplitudes presented in this work manifestly respect the large $p$ limit, as they should.
Note also that when $\tilde{s}=\tilde{t}=0$ one recovers \eqref{penedoneslim2}.

In this paper we have seen that an even more general version of flat-space limit seems to hold, which was noticed already in \cite{Aprile:2020mus} for $\mathcal{N}=4$ SYM correlators. In particular, all Mellin amplitudes can be written in terms of a pre-amplitude defined via \eqref{integral_tr_James}:
\be
\mathcal{M}_n = \frac{i}{2 \pi} \int_0^\infty  d\alpha \int_\mathcal{C} d\beta\ e^{-\alpha-\beta} \alpha^{\Sigma_p-1}  (-\beta)^{2-\Sigma_p} \, \tilde{\mathcal{M}}_n(S,T),
\ee
where
\be
S = \alpha s - \beta \tilde{s},\quad\tilde{S} = \alpha s + \beta \tilde{s}.
\ee
Quite nicely, the highest degree terms in $S,T$ are precisely the polynomials appearing in the expansion of the Veneziano amplitude.
In other words, the pre-amplitude $\tilde{\mathcal{M}}$ is given by
\begin{equation}
\tilde{\mathcal{M}}= \mathcal{V}_{\text{open}}(S,T)+\text{lower orders in}\,\, S,T,\tilde{S},\tilde{T}.
\end{equation}
Note that the sub-amplitude $\mathcal{V}_{\text{open}}(S,T)$ automatically contains the two flat-space limits just discussed, therefore in this sense it is a generalisation of those.

Finally, the covariantisation of the effective action suggests that the largest sub-amplitude related to flat space is the one obtained with the replacement of the partial derivatives with covariant ones \cite{Abl:2020dbx}, with all other limits just discussed arising as particular cases.
An explicit form for this sub-amplitude at all orders in $\alpha'$ is still not known, nor is the analogous expression for the AdS$_5 \times$S$^5$ background. We hope to report on this in the future.

\section{Outlook and conclusions}
\label{sec:conc}
In this paper we initiated the study of tree-level $\alpha'$ corrections to the four-point function of half-BPS operators in a $4d$, $\mathcal{N}=2$ SCFT with flavour group $SO(8)$, dual to type IIB string theory on AdS$_5 \times$S$^5/\mathbb{Z}_2$ \cite{Sen:1996vd,Fayyazuddin:1998fb,Aharony:1998xz}.
In particular, the strong-coupling expansion of these four-point correlators corresponds to the low-energy expansion of an AdS$_5 \times$S$^3$ version of the Veneziano amplitude. 
By generalising a procedure first proposed in \cite{Abl:2020dbx} in the context of $\mathcal{N}=4$ SYM, we conjectured that \emph{all} half-BPS four-point correlators can be obtained by evaluating generalised contact Witten diagrams whose vertices come from an $8$-dimensional effective potential written in terms of a single scalar field.
We then showed explicit results for the first four orders in $\alpha'$. 
We found that at each order in $\alpha'$ the various correlators are given by a {\it main} amplitude, which represents the covariantisation of the flat-space amplitude, plus a certain number of ambiguities which arise as a result of curvature effects of the background and therefore do not have a flat-space counterpart.
Nicely, the end results are remarkably simple when written in terms of an integral transform, which is perhaps the most natural generalisation of the integral proposed by Penedones in the flat-space limit \cite{Penedones:2010ue}.

While we believe that the simplicity of these correlators and the fact that they correctly capture the various flat-space limits present in literature are indicative of the validity of the method, the existence of the effective action is still to be proven and it would be interesting to find independent methods to check the conjecture or, more ambitiously, to derive it from first principles.
Another consistency check of our results could be provided by an analysis of the spectrum of anomalous dimensions. In AdS$_5 \times$S$^5$  \cite{Drummond:2020dwr,Aprile:2020mus}, this furnished an independent method that led to the same results as those of \cite{Abl:2020dbx}.
In particular, we expect the $\alpha'$-corrected anomalous dimensions of the exchanged double-trace operators to induce a splitting of the residual degeneracy left in the field-theory anomalous dimensions computed in \cite{Drummond:2022dxd}, as a result of the breaking of the hidden conformal symmetry \cite{Santagata:2022hga}.

We believe that this work can open a number of interesting directions.
\begin{itemize}
\item As mentioned already, these results, together with their AdS$_5 \times$S$^5$ analogous \cite{Aprile:2020mus,Abl:2020dbx}, suggest that there is a more general version of flat-space limit associated to the covariantisation of the derivatives in the effective action. Finding this sub-amplitude might shed light on new relations between flat and AdS amplitudes.
\item On the other hand, it is very important to find a systematic way to (list and) compute all the ambiguities, which represent true curvature effects. In AdS$_5 \times$S$^5$ they can be fixed with various techniques, such as localisation \cite{Binder:2019jwn,Chester:2019pvm,Chester:2020dja,Chester:2020vyz}, symmetry principles \cite{Abl:2020dbx}, dispersive sum rules \cite{Alday:2022uxp,Alday:2022xwz} or also bootstrap approaches \cite{Aprile:2020mus}. It would be interesting to see whether these methods can also applied to this background \cite{Behan:2023fqq}.
\item  With these new correlators at hand, it is now possible to extract novel CFT data. As we mentioned, we expect the $\alpha'$-corrections of the exchanged double-trace operators to drive a splitting of the residual degeneracy left in the field-theory anomalous dimensions.
The splitting of this degeneracy is controlled by a characteristic polynomial - an intrinsically non-perturbative object -  that enjoys a lot of intriguing features. It would be extremely interesting to find the form of this polynomial, perhaps also deriving the explicit dependence on the dimension $\theta_1,\theta_2$\footnote{In fact, it should also be the case that the supergravity anomalous dimensions in AdS$_3 \times$S$^3$ \cite{Aprile:2021mvq} undergo a similar splitting when string corrections are turned on \cite{Santagata:2022hga}.}.
\item It would also be interesting to compute higher genus $\alpha'$ corrections. We believe that similar results to those obtained in \cite{Aprile:2022tzr} for $\mathcal{N}=4$ SYM will find a natural generalisation to AdS$_5 \times$S$^3$. In particular, it should be the case that the accidental degeneracy enjoyed by certain classes of tree-level correlators gets broken at higher loops, through a phenomenon known as {\it sphere splitting} \cite{Aprile:2022tzr}.
\item Finally, as mentioned already, it is well known that open and closed string amplitudes are related by relations known as KLT relations \cite{Kawai:1985xq}; in addition, colour-ordered open string amplitudes are related each other through monodromy relations \cite{Stieberger:2009hq,Bjerrum-Bohr:2009ulz}. These are the uplifted ``stringy" versions of double-copy \cite{Bern:2010ue} of BCJ \cite{Bern:2008qj} relations, respectively, and reduce to the latter in the $\alpha' \rightarrow 0$ limit.
It is our belief that (a suitable generalisation of) these relations will also hold in this set-up, at least at the level of the main amplitude, since this object is directly connected to the flat amplitude.
From this point of view, it is promising that the generalised Mellin amplitudes in the field theory limit do satisfy double copy \cite{Zhou:2021gnu} and BCJ relations \cite{Drummond:2022dxd} completely analogous to flat space\footnote{For recent developments on double-copy and BCJ relations in different (A)dS backgrounds, see e.g.\cite{Farrow:2018yni,Lipstein:2019mpu,Armstrong:2020woi,Albayrak:2020fyp,Cheung:2022pdk,Herderschee:2022ntr,Lipstein:2023pih,Armstrong:2023phb} and references therein.} and moreover the highest degree terms in the pre-amplitude do enjoy these relations, since they literally coincide with the flat Veneziano amplitude written in terms of $S,T,U$ variables.
We hope to report on this in the near future.
\end{itemize}

\section*{Acknowledgement}
We thank James Drummond, Pietro Ferrero, Paul Heslop, Arthur Lipstein for providing important comment and feedback on the manuscript, James Drummond for collaboration at early stage of this project, Heng-Yu Chen and Yu-tin Huang for discussion on related topics.
MS is supported by Ministry of Science and Technology
(MOST) through the grant 110-2112-M-002-006-. RG is supported by the Mathematics and Theoretical Physics research group at the University of Hertfordshire.

\appendix

\section{Decorated Witten diagrams in Mellin space}
\label{app:Witten}
In this appendix we fill in the details between \eqref{wittengen} and \eqref{decoratemellin} for the derivation of the Mellin transform of the generalised Witten diagrams with decorations. The details follow closely the original presentation of \cite{Abl:2020dbx} and provide a slight generalisation to the case of AdS$_{\theta_1+1} \times$S$^{\theta_2+1}$ where the AdS and S dimensions no longer coincide.

Our starting point is the definition of the decorated AdS$\times$S Witten diagram which we repeat here for convenience
\begin{equation}
\prod_i \mathcal{C}_{\Delta_i} \frac{\prod_{i<j} (X_i \cdot X_j)^{n_{ij}^X}(Y_i \cdot Y_j)^{n_{ij}^Y}}{(-2)^{2\Sigma_\Delta}}\int_{\text{AdS} \times \text{S}} d^{\theta_1+1}\hat{X} d^{\theta_2+1} \hat{Y}\prod_{i=1}^4 \frac{P_i^{n_i^P} Q_i^{n_i^Q}(\Delta_i)_{n_i}}{(P_i+Q_i)^{\Delta_i+n_i}}.
\label{eq:decorated_witten}
\end{equation}
The propogators in the denominator may be Taylor-expanded using 
\begin{equation}
\frac{1}{(P_i+Q_i)^{\Delta_i + n_i}} = \sum_{p_i=0}^{\infty} (-1)^{p_i} \frac{(p_i+1)_{\Delta_i + n_i-1}}{\Gamma(\Delta_i + n_i)}\frac{Q_i^{p_i}}{P_i^{p_i+\Delta_i + n_i}},
\end{equation}
from which we arrive at an expansion for the decorated $AdS\times S$ Witten diagram in terms of regular AdS diagrams and their sphere counterparts given explicitly by
\begin{equation}
\prod_i \mathcal{C}_{\Delta_i}(-2)^{2\Sigma_X+2\Sigma_Y}\left(\prod_{i<j} (X_i \cdot X_j)^{n_{ij}^X}(Y_i \cdot Y_j)^{n_{ij}^Y}\right)\sum_{p_i=0}^{\infty}\prod_{i=1}^4 (-)^{p_i} \frac{(p_i+1)_{\Delta_i+n_i-1}}{\Gamma(\Delta_i)} D^{(\theta_1)}_{\Delta_i+ p_i  +n_i -n_i^P} B^{(\theta_2)}_{p_i+n_i^Q}.
\end{equation}
Substituting in the following expressions 
\begin{align}
D^{(\theta_1)}_{\Delta_i+p_i+n_i-n_i^P} &= \frac{\frac{1}{2} \pi^{\theta_1/2}\Gamma(\Sigma_{\Delta}+\Sigma_{p}+\Sigma_{Q}+\Sigma_{X}+\Sigma_{Y}-\theta_1/2)}{(-2)^{\Sigma_{\Delta}+\Sigma_{p}+\Sigma_{Q}+\Sigma_{X}+\Sigma_{Y}}\prod_i \Gamma(\Delta_i+p_i+n_i-n_i^P)} \int \frac{d \delta_{ij}}{(2\pi i)^2} \prod_{i<j} \frac{\Gamma(\delta_{ij})}{(X_i \cdot X_j)^{\delta_{ij}}}, \notag \\
B^{(\theta_2)}_{p_i+n_i^Q} &= 2 \cdot 2^{\Sigma_p+\Sigma_Q} \frac{\pi^{\theta_2 /2 +1} \prod_i \Gamma(p_i+n_i^Q+1)}{\Gamma(\Sigma_p+\Sigma_Q + \theta_2/2+1)} \sum_{\{ d_{ij}\}}\prod_{i<j} \frac{(Y_i \cdot Y_j)^{d_{ij}}}{\Gamma(d_{ij}+1)},
\end{align}
for which we have
\begin{equation}
 \sum_i \delta_{ij}=\Delta_j+p_j+n_j-n_j^P, \quad \sum_i d_{ij}=p_j+n_j^Q,
\end{equation}
the decorated Witten diagram \eqref{eq:decorated_witten} is found to take the following form
\begin{align}
& \frac{\pi^{\frac{\theta_1+\theta_2}{2}}}{(-2)^{\Sigma_{\Delta}}}\left(\prod_i \frac{\mathcal{C}_{\Delta_i}}{\Gamma(\Delta_i)} \right)\sum_{p_i=0}^{\infty}(-)^{ \Sigma_{p}} \int \frac{d \delta_{ij}}{(2\pi i)^2} \sum_{\{ d_{ij}\}} \prod_{i<j} \left(\frac{(Y_i \cdot Y_j)^{d_{ij}}}{(X_i \cdot X_j)^{\delta_{ij}}}\frac{\Gamma(\delta_{ij})}{\Gamma(d_{ij}+1)}\right) \mathcal{M}_{\Delta_i}[(\ref{eq:decorated_witten})].
\end{align}
Here we have introduced the notation $\Sigma_Q$ for half the sum over the $n_i^Q$ and $\Sigma_X$, $\Sigma_Y$ for the sum over the $n_{ij}^X$, $n_{ij}^Y$, respectively.
Here $\mathcal{M}_{\Delta_i}[(\ref{eq:decorated_witten})]$ is the desired representation of \eqref{eq:decorated_witten} in generalised Mellin space and is given by
\begin{align}
\mathcal{M}_{\Delta_i}[(\ref{eq:decorated_witten})] =& (-2)^{\Sigma_X}(2)^{\Sigma_Y} (-)^{2\Sigma_{Q}}  \left(\prod_{i=1}^4 \left( p_i+n_i^X+\Delta_i \right)_{n_i^P} ( p_i-n_i^Q-n_i^Y+1 )_{n_i^Q} \right)  \times \notag\\
&  \left( \prod_{i<j}(\delta_{ij})_{n_{i,j}^X} (d_{ij}-n^Y_{ij}+1)_{n_{ij}^Y}\right)(\Sigma _{p}-\Sigma _Y+\frac{\theta _2}{2}+1)_{\Sigma _{\Delta }-\frac{\theta _1+\theta _2}{2}-1 +\Sigma _X+\Sigma _Y},
\end{align}
where we have the contraints $ \sum_i d_{ij}=p_j$ and $\sum_i \delta_{ij} =\Delta_j +p_j$.

\section{Details at $\alpha'^5$}
\label{app:details}
In this appendix we collect some explicit expression for the amplitudes at $\alpha'^5$.
First, the computation at this order requires the evaluation of $\nabla_\mu \nabla_\nu \nabla_\rho \mathcal{W}_1^3 \,\, \nabla_\mu \nabla_\nu \nabla_\rho \mathcal{W}_2^3$ and $\nabla^2 \nabla_\mu \nabla_\nu \mathcal{W}_1^3 \,\, \nabla_\nu \nabla_\mu \mathcal{W}_2^3$. 
We find
\begin{equation}
\begin{split}
& \nabla_\mu \nabla_\nu \nabla_\rho \mathcal{W}_1^3 \,\, \nabla_\mu \nabla_\nu \nabla_\rho \mathcal{W}_2^3 = \frac{9}{\mathcal{W}_1^6 \mathcal{W}_2^6}\biggl(245 P_1^3 P_2^3+62 P_1^2 P_2^3 Q_1-3 P_1 P_2^3 Q_1^2+62 P_1^3 P_2^2 Q_2 \notag \\
&-575 P_1^2 P_2^2 Q_1 Q_2-100 P_1 P_2^2 Q_1^2 Q_2-3 P_2^2 Q_1^3 Q_2-3 P_1^3 P_2 Q_2^2-100 P_1^2 P_2 Q_1 Q_2^2+457 P_1 P_2 Q_1^2 Q_2^2 \notag \\
&+14 P_2 Q_1^3 Q_2^2-3 P_1^2 Q_1 Q_2^3+14 P_1 Q_1^2 Q_2^3-163 Q_1^3 Q_2^3+1045 P_1^2 P_2^2 (X_1 \cdot X_2)+62 P_1 P_2^2 Q_1 (X_1 \cdot X_2) \notag \\
&-3 P_2^2 Q_1^2 (X_1 \cdot X_2)+62 P_1^2 P_2 Q_2 (X_1 \cdot X_2)-2012 P_1 P_2 Q_1 Q_2 (X_1 \cdot X_2)-114 P_2 Q_1^2 Q_2 (X_1 \cdot X_2) \notag \\
&-3 P_1^2 Q_2^2 (X_1 \cdot X_2)-114 P_1 Q_1 Q_2^2 (X_1 \cdot X_2)+869 Q_1^2 Q_2^2 (X_1 \cdot X_2)+1200 P_1 P_2 (X_1 \cdot X_2)^2 \notag \\
&-1200 Q_1 Q_2 (X_1 \cdot X_2)^2+400 (X_1 \cdot X_2)^3+963 P_1^2 P_2^2 (Y_1 \cdot Y_2)-14 P_1 P_2^2 Q_1 (Y_1 \cdot Y_2)+3 P_2^2 Q_1^2 (Y_1 \cdot Y_2) \notag \\
&-14 P_1^2 P_2 Q_2 (Y_1 \cdot Y_2)-1988 P_1 P_2 Q_1 Q_2 (Y_1 \cdot Y_2)-14 P_2 Q_1^2 Q_2 (Y_1 \cdot Y_2)+3 P_1^2 Q_2^2 (Y_1 \cdot Y_2) \notag \\
&-14 P_1 Q_1 Q_2^2 (Y_1 \cdot Y_2)+963 Q_1^2 Q_2^2 (Y_1 \cdot Y_2)+2400 P_1 P_2 (X_1 \cdot X_2) (Y_1 \cdot Y_2)-2400 Q_1 Q_2 (X_1 \cdot X_2) (Y_1 \cdot Y_2) \notag \\
&+1200 (X_1 \cdot X_2)^2 (Y_1 \cdot Y_2)+1200 P_1 P_2 (Y_1 \cdot Y_2)^2-1200 Q_1 Q_2 (Y_1 \cdot Y_2)^2+1200 (X_1 \cdot X_2) (Y_1 \cdot Y_2)^2 \notag \\
&+400 (Y_1 \cdot Y_2)^3\biggr) \equiv \frac{9}{\mathcal{W}_1^6 \mathcal{W}_2^6} Z_{12}  \\
\end{split}
\end{equation}
and
\begin{equation}
\begin{split}
& \nabla^2 \nabla_\mu \nabla_\nu \mathcal{W}_1^3 \,\, \nabla_\nu \nabla_\mu \mathcal{W}_2^3 = \frac{9}{\mathcal{W}_1^5\mathcal{W}_2^5}  \biggl( 31Q_{1}^{2}Q_{2}^{2} -167P_{1}^{2}P_{2}^{2} - 3P_{1}P_{2}^2 Q_{1} - 3P_{1}^2P_{2} Q_{2} +3P_{1}Q_{1}Q_{2}^{2} \\
& + 3P_{2}Q_{1}^{2}Q_{2}+ 136P_{1}P_{2}Q_{1}Q_{2}   - 416P_{1}P_{2}(X_{1}\cdot X_{2}) + 160Q_{1}Q_{2}(X_{1} \cdot X_{2})  - 160(X_{1} \cdot X_{2})(Y_{1} \cdot Y_{2})  \\
&-160P_{1}P_{2}(Y_{1} \cdot Y_{2}) -96 Q_{1}Q_{2}(Y_{1} \cdot Y_{2})  - 208(X_{1}\cdot X_{2})^{2}+ 48(Y_{1} \cdot Y_{2})^{2} \biggr).
\end{split}
\end{equation}
Then, the main amplitude is given by the following integral
\begin{equation}
\begin{split}
 & \langle \mathcal{O}\mathcal{O}\mathcal{O}\mathcal{O} \rangle^{\text{main}} |_{\alpha'^5} (1234)   = \frac{1}{3} \frac{1}{32}  \frac{3^2 \,\mathcal{C}_3^4}{(-2)^{12}} \int_{\text{AdS}_5 \times\text{S}^3} d^5 \hat{X} d^3 \hat{Y} \frac{1}{(\mathcal{W}_1)^3(\mathcal{W}_2)^3(\mathcal{W}_3)^3(\mathcal{W}_4)^3} \\
& \left[  \left(\frac{1}{6}\zeta_3 \pi^2+\zeta_5\right)2 \left( \frac{Z_{12}}{(\mathcal{W}_1)^3(\mathcal{W}_2)^3}+ \frac{Z_{34}}{(\mathcal{W}_3)^3(\mathcal{W}_4)^3}+ \frac{Z_{14}}{(\mathcal{W}_1)^3(\mathcal{W}_4)^3}+ \frac{Z_{23}}{(\mathcal{W}_2)^3(\mathcal{W}_3)^3} \right) \right.+ \\
&\left. \left(\frac{1}{6}\zeta_3 \pi^2-2\zeta_5\right)2 \left( \frac{Z_{13}}{(\mathcal{W}_1)^3(\mathcal{W}_3)^3}+ \frac{Z_{24}}{(\mathcal{W}_2)^3(\mathcal{W}_4)^3}\right) \right].
\end{split}
\end{equation}
The associated generalised Mellin amplitude reads
\begin{equation}
{\cal M}_3^{\text{main}}=\frac{1}{3} \left(\frac{1}{6}\zeta_3 \pi^2+\zeta_5\right) ({\cal M}_3^s+{\cal M}_3^t) +\frac{1}{3}\left(\frac{1}{6}\zeta_3 \pi^2-2\zeta_5\right) 
 {\cal M}_3^u,
\end{equation}
where
\begin{align}
{\cal M}_3^s=\, & (\Sigma_p-2)_5 {\bf s}^3 - \frac{3}{2}\Big[10 \tilde s + 2 \Sigma_p -1 \Big] (\Sigma_p-2)_4{\bf s}^2 \notag\\
& + \frac{1}{8}\Big[ 480 \tilde{s}^2 + 120(1+\Sigma_p)\tilde{s} +16\Sigma_p(2+\Sigma_p)-11P+142\Big](\Sigma_p-2)_3{\bf s} \notag  \\
&  -\frac{1}{16}\Big[ 480(2 \tilde{s}+3) \tilde{s}^2 - 2(33P-164 \Sigma_p-654)\tilde{s} \notag \\
& \quad \quad  \quad \quad\quad\quad -16(p_1 p_2 +p_3 p_4)+4 \Sigma_p (25+8\Sigma_p)-33P \Big] (\Sigma_p-2)_2.
\end{align}
Similarly, the $s$-channel ambiguity is given by
\begin{align}
{\cal M}_{\text{3,amb}}^s&=(\Sigma_p-2)_4 {\bf s}^2 -\frac{1}{13}\Big[ 8(7+2 \Sigma_p)\tilde{s} +\Sigma_p(5+8\Sigma_p)-4(p_1 p_2 +p_3 p_4)-19 \Big](\Sigma_p-2)_3 {\bf s} \notag \\
& +\frac{1}{104}\Big [ 32(16\Sigma_p -9)\tilde{s}^2 + 32(10 \Sigma_p^2 - 8 \Sigma -3p_1 p_2 - 3p_3 p_4 -15) \tilde{s} \notag \\
&\quad \quad  \quad \quad \quad \quad \quad \quad \quad \quad \quad+16(p_1+p_2)(p_3+p_4)(2\Sigma-1)  +7P-28 \Sigma_p\Big] (\Sigma_p-2)_2.
\end{align}
The expressions for the related pre-amplitudes are given in the main body, see \eqref{alpha5main} and \eqref{alpha5amb} for the Mellin pre-amplitudes $\tilde{{\cal M}}_3^s$, $\tilde{\cal M}_{\text{3,amb}}^s$, respectively.

All in all, the full colour-ordered Mellin amplitude (including all other ambiguities from previous orders) reads
\begin{equation}
\begin{split}
{\cal M}_3(1234) = &  \frac{1}{3} \left(\frac{1}{6}\zeta_3 \pi^2+\zeta_5\right) ({\cal M}_3^s+{\cal M}_3^t) + \frac{1}{3}\left(\frac{1}{6}\zeta_3 \pi^2-2\zeta_5\right) 
 {\cal M}_3^u + \\
 &+ a_3 + b_2({\cal M}_1^s+{\cal M}_1^t)+ c_1({ \cal M}_2^s+ {\cal M}_2^t)+d_1 {\cal M}_2^u+ e_2 ({\cal M}_{2,\text{amb}}^s+{\cal M}_{2,\text{amb}}^t) +\\
& + f_2 {\cal M}_{2,\text{amb}}^u + l_1({\cal M}_{\text{3,amb}}^s+{\cal M}_{\text{3,amb}}^t)+ h_1{\cal M}_{3,\text{amb}}^u.
\end{split}
\end{equation}

\bibliographystyle{JHEP}
\bibliography{new_bib}

\providecommand{\href}[2]{#2}\begingroup\raggedright\begin{thebibliography}{10}

\bibitem{Maldacena:1997re}
J.~M. Maldacena, \emph{{The Large N limit of superconformal field theories and
  supergravity}}, \href{https://doi.org/10.4310/ATMP.1998.v2.n2.a1}{\emph{Adv.
  Theor. Math. Phys.} {\bfseries 2} (1998) 231}
  [\href{https://arxiv.org/abs/hep-th/9711200}{{\ttfamily hep-th/9711200}}].

\bibitem{Rastelli:2017udc}
L.~Rastelli and X.~Zhou, \emph{{How to Succeed at Holographic Correlators
  Without Really Trying}},
  \href{https://doi.org/10.1007/JHEP04(2018)014}{\emph{JHEP} {\bfseries 04}
  (2018) 014} [\href{https://arxiv.org/abs/1710.05923}{{\ttfamily
  1710.05923}}].

\bibitem{Rastelli:2016nze}
L.~Rastelli and X.~Zhou, \emph{{Mellin amplitudes for $AdS_5\times S^5$}},
  \href{https://doi.org/10.1103/PhysRevLett.118.091602}{\emph{Phys. Rev. Lett.}
  {\bfseries 118} (2017) 091602}
  [\href{https://arxiv.org/abs/1608.06624}{{\ttfamily 1608.06624}}].

\bibitem{Alday:2018pdi}
L.~F. Alday, A.~Bissi and E.~Perlmutter, \emph{{Genus-One String Amplitudes
  from Conformal Field Theory}},
  \href{https://doi.org/10.1007/JHEP06(2019)010}{\emph{JHEP} {\bfseries 06}
  (2019) 010} [\href{https://arxiv.org/abs/1809.10670}{{\ttfamily
  1809.10670}}].

\bibitem{Alday:2018kkw}
L.~F. Alday, \emph{{On genus-one string amplitudes on $AdS_5 \times S^5$}},
  \href{https://doi.org/10.1007/JHEP04(2021)005}{\emph{JHEP} {\bfseries 04}
  (2021) 005} [\href{https://arxiv.org/abs/1812.11783}{{\ttfamily
  1812.11783}}].

\bibitem{Alday:2019nin}
L.~F. Alday and X.~Zhou, \emph{{Simplicity of AdS Supergravity at One Loop}},
  \href{https://doi.org/10.1007/JHEP09(2020)008}{\emph{JHEP} {\bfseries 09}
  (2020) 008} [\href{https://arxiv.org/abs/1912.02663}{{\ttfamily
  1912.02663}}].

\bibitem{Aprile:2017qoy}
F.~Aprile, J.~M. Drummond, P.~Heslop and H.~Paul, \emph{{Loop corrections for
  Kaluza-Klein AdS amplitudes}},
  \href{https://doi.org/10.1007/JHEP05(2018)056}{\emph{JHEP} {\bfseries 05}
  (2018) 056} [\href{https://arxiv.org/abs/1711.03903}{{\ttfamily
  1711.03903}}].

\bibitem{Aprile:2017bgs}
F.~Aprile, J.~M. Drummond, P.~Heslop and H.~Paul, \emph{{Quantum Gravity from
  Conformal Field Theory}},
  \href{https://doi.org/10.1007/JHEP01(2018)035}{\emph{JHEP} {\bfseries 01}
  (2018) 035} [\href{https://arxiv.org/abs/1706.02822}{{\ttfamily
  1706.02822}}].

\bibitem{Aprile:2019rep}
F.~Aprile, J.~Drummond, P.~Heslop and H.~Paul, \emph{{One-loop amplitudes in
  AdS$_{5} \times S^{5}$ supergravity from $ \mathcal{N} $ = 4 SYM at strong
  coupling}}, \href{https://doi.org/10.1007/JHEP03(2020)190}{\emph{JHEP}
  {\bfseries 03} (2020) 190}
  [\href{https://arxiv.org/abs/1912.01047}{{\ttfamily 1912.01047}}].

\bibitem{Drummond:2019odu}
J.~M. Drummond, D.~Nandan, H.~Paul and K.~S. Rigatos, \emph{{String corrections
  to AdS amplitudes and the double-trace spectrum of $ \mathcal{N} $ = 4 SYM}},
  \href{https://doi.org/10.1007/JHEP12(2019)173}{\emph{JHEP} {\bfseries 12}
  (2019) 173} [\href{https://arxiv.org/abs/1907.00992}{{\ttfamily
  1907.00992}}].

\bibitem{Drummond:2019hel}
J.~M. Drummond and H.~Paul, \emph{{One-loop string corrections to AdS
  amplitudes from CFT}},
  \href{https://doi.org/10.1007/JHEP03(2021)038}{\emph{JHEP} {\bfseries 03}
  (2021) 038} [\href{https://arxiv.org/abs/1912.07632}{{\ttfamily
  1912.07632}}].

\bibitem{Goncalves:2014ffa}
V.~Gon\c{c}alves, \emph{{Four point function of $\mathcal{N}=4$ stress-tensor
  multiplet at strong coupling}},
  \href{https://doi.org/10.1007/JHEP04(2015)150}{\emph{JHEP} {\bfseries 04}
  (2015) 150} [\href{https://arxiv.org/abs/1411.1675}{{\ttfamily 1411.1675}}].

\bibitem{Drummond:2020dwr}
J.~M. Drummond, H.~Paul and M.~Santagata, \emph{{Bootstrapping string theory on
  AdS$_5 \times S^5$}},  \href{https://arxiv.org/abs/2004.07282}{{\ttfamily
  2004.07282}}.

\bibitem{Aprile:2020mus}
F.~Aprile, J.~M. Drummond, H.~Paul and M.~Santagata, \emph{{The
  Virasoro-Shapiro amplitude in AdS$_{5} \times S^{5}$ and level splitting of
  10d conformal symmetry}},
  \href{https://doi.org/10.1007/JHEP11(2021)109}{\emph{JHEP} {\bfseries 11}
  (2021) 109} [\href{https://arxiv.org/abs/2012.12092}{{\ttfamily
  2012.12092}}].

\bibitem{Bissi:2020wtv}
A.~Bissi, G.~Fardelli and A.~Georgoudis, \emph{{Towards all loop supergravity
  amplitudes on AdS$_5 \times$ S$^5$}},
  \href{https://doi.org/10.1103/PhysRevD.104.L041901}{\emph{Phys. Rev. D}
  {\bfseries 104} (2021) L041901}
  [\href{https://arxiv.org/abs/2002.04604}{{\ttfamily 2002.04604}}].

\bibitem{Bissi:2020woe}
A.~Bissi, G.~Fardelli and A.~Georgoudis, \emph{{All loop structures in
  supergravity amplitudes on AdS$_5 \times$ S$^5$ from CFT}},
  \href{https://doi.org/10.1088/1751-8121/ac0ebf}{\emph{J. Phys. A} {\bfseries
  54} (2021) 324002} [\href{https://arxiv.org/abs/2010.12557}{{\ttfamily
  2010.12557}}].

\bibitem{Aprile:2020luw}
F.~Aprile and P.~Vieira, \emph{{Large $p$ explorations. From SUGRA to big
  STRINGS in Mellin space}},
  \href{https://doi.org/10.1007/JHEP12(2020)206}{\emph{JHEP} {\bfseries 12}
  (2020) 206} [\href{https://arxiv.org/abs/2007.09176}{{\ttfamily
  2007.09176}}].

\bibitem{Drummond:2020uni}
J.~M. Drummond, R.~Glew and H.~Paul, \emph{{One-loop string corrections for AdS
  Kaluza-Klein amplitudes}},
  \href{https://doi.org/10.1007/JHEP12(2021)072}{\emph{JHEP} {\bfseries 12}
  (2021) 072} [\href{https://arxiv.org/abs/2008.01109}{{\ttfamily
  2008.01109}}].

\bibitem{Aprile:2022tzr}
F.~Aprile, J.~M. Drummond, R.~Glew and M.~Santagata, \emph{{One-loop string
  amplitudes in AdS$_{5} \times S^{5}$: Mellin space and sphere splitting}},
  \href{https://doi.org/10.1007/JHEP02(2023)087}{\emph{JHEP} {\bfseries 02}
  (2023) 087} [\href{https://arxiv.org/abs/2207.13084}{{\ttfamily
  2207.13084}}].

\bibitem{Huang:2021xws}
Z.~Huang and E.~Y. Yuan, \emph{{Graviton scattering in AdS$_5 \times$ S$^5$ at
  two loops}}, \href{https://doi.org/10.1007/JHEP04(2023)064}{\emph{JHEP}
  {\bfseries 04} (2023) 064}
  [\href{https://arxiv.org/abs/2112.15174}{{\ttfamily 2112.15174}}].

\bibitem{Drummond:2022dxw}
J.~M. Drummond and H.~Paul, \emph{{Two-loop supergravity on AdS$_5 \times$
  S$^5$ from CFT}}, \href{https://doi.org/10.1007/JHEP08(2022)275}{\emph{JHEP}
  {\bfseries 08} (2022) 275}
  [\href{https://arxiv.org/abs/2204.01829}{{\ttfamily 2204.01829}}].

\bibitem{Alday:2022uxp}
L.~F. Alday, T.~Hansen and J.~A. Silva, \emph{{AdS Virasoro-Shapiro from
  dispersive sum rules}},
  \href{https://doi.org/10.1007/JHEP10(2022)036}{\emph{JHEP} {\bfseries 10}
  (2022) 036} [\href{https://arxiv.org/abs/2204.07542}{{\ttfamily
  2204.07542}}].

\bibitem{Alday:2022xwz}
L.~F. Alday, T.~Hansen and J.~A. Silva, \emph{{AdS Virasoro-Shapiro from
  single-valued periods}},
  \href{https://doi.org/10.1007/JHEP12(2022)010}{\emph{JHEP} {\bfseries 12}
  (2022) 010} [\href{https://arxiv.org/abs/2209.06223}{{\ttfamily
  2209.06223}}].

\bibitem{Heslop:2022xgp}
P.~Heslop, \emph{{The SAGEX Review on Scattering Amplitudes, Chapter 8: Half
  BPS correlators}}, \href{https://doi.org/10.1088/1751-8121/ac8c71}{\emph{J.
  Phys. A} {\bfseries 55} (2022) 443009}
  [\href{https://arxiv.org/abs/2203.13019}{{\ttfamily 2203.13019}}].

\bibitem{Bissi:2022mrs}
A.~Bissi, A.~Sinha and X.~Zhou, \emph{{Selected topics in analytic conformal
  bootstrap: A guided journey}},
  \href{https://doi.org/10.1016/j.physrep.2022.09.004}{\emph{Phys. Rept.}
  {\bfseries 991} (2022) 1} [\href{https://arxiv.org/abs/2202.08475}{{\ttfamily
  2202.08475}}].

\bibitem{Caron-Huot:2018kta}
S.~Caron-Huot and A.-K. Trinh, \emph{{All tree-level correlators in AdS$_{5}
  \times S_{5}$ supergravity: hidden ten-dimensional conformal symmetry}},
  \href{https://doi.org/10.1007/JHEP01(2019)196}{\emph{JHEP} {\bfseries 01}
  (2019) 196} [\href{https://arxiv.org/abs/1809.09173}{{\ttfamily
  1809.09173}}].

\bibitem{Caron-Huot:2021usw}
S.~Caron-Huot and F.~Coronado, \emph{{Ten dimensional symmetry of $ \mathcal{N}
  $ = 4 SYM correlators}},
  \href{https://doi.org/10.1007/JHEP03(2022)151}{\emph{JHEP} {\bfseries 03}
  (2022) 151} [\href{https://arxiv.org/abs/2106.03892}{{\ttfamily
  2106.03892}}].

\bibitem{Caron-Huot:2023wdh}
S.~Caron-Huot, F.~Coronado and B.~M\"uhlmann, \emph{{Determinants in self-dual
  N=4 SYM and twistor space}},
  \href{https://arxiv.org/abs/2304.12341}{{\ttfamily 2304.12341}}.

\bibitem{Abl:2020dbx}
T.~Abl, P.~Heslop and A.~E. Lipstein, \emph{{Towards the Virasoro-Shapiro
  amplitude in AdS$_{5} \times S^{5}$}},
  \href{https://doi.org/10.1007/JHEP04(2021)237}{\emph{JHEP} {\bfseries 04}
  (2021) 237} [\href{https://arxiv.org/abs/2012.12091}{{\ttfamily
  2012.12091}}].

\bibitem{Abl:2021mxo}
T.~Abl, P.~Heslop and A.~E. Lipstein, \emph{{Higher-dimensional symmetry of
  AdS$_{2} \times S^{2}$ correlators}},
  \href{https://doi.org/10.1007/JHEP03(2022)076}{\emph{JHEP} {\bfseries 03}
  (2022) 076} [\href{https://arxiv.org/abs/2112.09597}{{\ttfamily
  2112.09597}}].

\bibitem{Giusto:2019pxc}
S.~Giusto, R.~Russo, A.~Tyukov and C.~Wen, \emph{{Holographic correlators in
  AdS$_3$ without Witten diagrams}},
  \href{https://doi.org/10.1007/JHEP09(2019)030}{\emph{JHEP} {\bfseries 09}
  (2019) 030} [\href{https://arxiv.org/abs/1905.12314}{{\ttfamily
  1905.12314}}].

\bibitem{Rastelli:2019gtj}
L.~Rastelli, K.~Roumpedakis and X.~Zhou, \emph{{$\mathbf{AdS_3\times S^3}$
  Tree-Level Correlators: Hidden Six-Dimensional Conformal Symmetry}},
  \href{https://doi.org/10.1007/JHEP10(2019)140}{\emph{JHEP} {\bfseries 10}
  (2019) 140} [\href{https://arxiv.org/abs/1905.11983}{{\ttfamily
  1905.11983}}].

\bibitem{Giusto:2020neo}
S.~Giusto, R.~Russo, A.~Tyukov and C.~Wen, \emph{{The CFT$_6$ origin of all
  tree-level 4-point correlators in AdS$_3 \times S^3$}},
  \href{https://doi.org/10.1140/epjc/s10052-020-8300-4}{\emph{Eur. Phys. J. C}
  {\bfseries 80} (2020) 736}
  [\href{https://arxiv.org/abs/2005.08560}{{\ttfamily 2005.08560}}].

\bibitem{Sen:1996vd}
A.~Sen, \emph{{F theory and orientifolds}},
  \href{https://doi.org/10.1016/0550-3213(96)00347-1}{\emph{Nucl. Phys. B}
  {\bfseries 475} (1996) 562}
  [\href{https://arxiv.org/abs/hep-th/9605150}{{\ttfamily hep-th/9605150}}].

\bibitem{Fayyazuddin:1998fb}
A.~Fayyazuddin and M.~Spalinski, \emph{{Large N superconformal gauge theories
  and supergravity orientifolds}},
  \href{https://doi.org/10.1016/S0550-3213(98)00545-8}{\emph{Nucl. Phys. B}
  {\bfseries 535} (1998) 219}
  [\href{https://arxiv.org/abs/hep-th/9805096}{{\ttfamily hep-th/9805096}}].

\bibitem{Aharony:1998xz}
O.~Aharony, A.~Fayyazuddin and J.~M. Maldacena, \emph{{The Large N limit of
  N=2, N=1 field theories from three-branes in F theory}},
  \href{https://doi.org/10.1088/1126-6708/1998/07/013}{\emph{JHEP} {\bfseries
  07} (1998) 013} [\href{https://arxiv.org/abs/hep-th/9806159}{{\ttfamily
  hep-th/9806159}}].

\bibitem{Alday:2021odx}
L.~F. Alday, C.~Behan, P.~Ferrero and X.~Zhou, \emph{{Gluon Scattering in AdS
  from CFT}}, \href{https://doi.org/10.1007/JHEP06(2021)020}{\emph{JHEP}
  {\bfseries 06} (2021) 020}
  [\href{https://arxiv.org/abs/2103.15830}{{\ttfamily 2103.15830}}].

\bibitem{Alday:2021ajh}
L.~F. Alday, A.~Bissi and X.~Zhou, \emph{{One-loop gluon amplitudes in AdS}},
  \href{https://doi.org/10.1007/JHEP02(2022)105}{\emph{JHEP} {\bfseries 02}
  (2022) 105} [\href{https://arxiv.org/abs/2110.09861}{{\ttfamily
  2110.09861}}].

\bibitem{Huang:2023oxf}
Z.~Huang, B.~Wang, E.~Y. Yuan and X.~Zhou, \emph{{AdS super gluon scattering up
  to two loops: A position space approach}},
  \href{https://arxiv.org/abs/2301.13240}{{\ttfamily 2301.13240}}.

\bibitem{Drummond:2022dxd}
J.~M. Drummond, R.~Glew and M.~Santagata, \emph{{BCJ relations in ${AdS}_5
  \times S^3$ and the double-trace spectrum of super gluons}},
  \href{https://arxiv.org/abs/2202.09837}{{\ttfamily 2202.09837}}.

\bibitem{Penedones:2010ue}
J.~Penedones, \emph{{Writing CFT correlation functions as AdS scattering
  amplitudes}}, \href{https://doi.org/10.1007/JHEP03(2011)025}{\emph{JHEP}
  {\bfseries 03} (2011) 025} [\href{https://arxiv.org/abs/1011.1485}{{\ttfamily
  1011.1485}}].

\bibitem{Fitzpatrick:2011hu}
A.~L. Fitzpatrick and J.~Kaplan, \emph{{Analyticity and the Holographic
  S-Matrix}}, \href{https://doi.org/10.1007/JHEP10(2012)127}{\emph{JHEP}
  {\bfseries 10} (2012) 127} [\href{https://arxiv.org/abs/1111.6972}{{\ttfamily
  1111.6972}}].

\bibitem{Behan:2023fqq}
C.~Behan, S.~M. Chester and P.~Ferrero, \emph{{Gluon scattering in AdS at
  finite string coupling from localization}},
  \href{https://arxiv.org/abs/2305.01016}{{\ttfamily 2305.01016}}.

\bibitem{Karch:2002sh}
A.~Karch and E.~Katz, \emph{{Adding flavor to AdS / CFT}},
  \href{https://doi.org/10.1088/1126-6708/2002/06/043}{\emph{JHEP} {\bfseries
  06} (2002) 043} [\href{https://arxiv.org/abs/hep-th/0205236}{{\ttfamily
  hep-th/0205236}}].

\bibitem{Nirschl:2004pa}
M.~Nirschl and H.~Osborn, \emph{{Superconformal Ward identities and their
  solution}},
  \href{https://doi.org/10.1016/j.nuclphysb.2005.01.013}{\emph{Nucl. Phys. B}
  {\bfseries 711} (2005) 409}
  [\href{https://arxiv.org/abs/hep-th/0407060}{{\ttfamily hep-th/0407060}}].

\bibitem{Schlotterer:2012ny}
O.~Schlotterer and S.~Stieberger, \emph{{Motivic Multiple Zeta Values and
  Superstring Amplitudes}},
  \href{https://doi.org/10.1088/1751-8113/46/47/475401}{\emph{J. Phys. A}
  {\bfseries 46} (2013) 475401}
  [\href{https://arxiv.org/abs/1205.1516}{{\ttfamily 1205.1516}}].

\bibitem{Kawai:1985xq}
H.~Kawai, D.~C. Lewellen and S.~H.~H. Tye, \emph{{A Relation Between Tree
  Amplitudes of Closed and Open Strings}},
  \href{https://doi.org/10.1016/0550-3213(86)90362-7}{\emph{Nucl. Phys. B}
  {\bfseries 269} (1986) 1}.

\bibitem{Stieberger:2009hq}
S.~Stieberger, \emph{{Open \& Closed vs. Pure Open String Disk Amplitudes}},
  \href{https://arxiv.org/abs/0907.2211}{{\ttfamily 0907.2211}}.

\bibitem{Bjerrum-Bohr:2009ulz}
N.~E.~J. Bjerrum-Bohr, P.~H. Damgaard and P.~Vanhove, \emph{{Minimal Basis for
  Gauge Theory Amplitudes}},
  \href{https://doi.org/10.1103/PhysRevLett.103.161602}{\emph{Phys. Rev. Lett.}
  {\bfseries 103} (2009) 161602}
  [\href{https://arxiv.org/abs/0907.1425}{{\ttfamily 0907.1425}}].

\bibitem{Bern:2008qj}
Z.~Bern, J.~J.~M. Carrasco and H.~Johansson, \emph{{New Relations for
  Gauge-Theory Amplitudes}},
  \href{https://doi.org/10.1103/PhysRevD.78.085011}{\emph{Phys. Rev. D}
  {\bfseries 78} (2008) 085011}
  [\href{https://arxiv.org/abs/0805.3993}{{\ttfamily 0805.3993}}].

\bibitem{Sleight:2016hyl}
C.~Sleight, \emph{{Interactions in Higher-Spin Gravity: a Holographic
  Perspective}}, \href{https://doi.org/10.1088/1751-8121/aa820c}{\emph{J. Phys.
  A} {\bfseries 50} (2017) 383001}
  [\href{https://arxiv.org/abs/1610.01318}{{\ttfamily 1610.01318}}].

\bibitem{Chen:2020ipe}
H.-Y. Chen and J.-i. Sakamoto, \emph{{Superconformal Block from Holographic
  Geometry}}, \href{https://doi.org/10.1007/JHEP07(2020)028}{\emph{JHEP}
  {\bfseries 07} (2020) 028}
  [\href{https://arxiv.org/abs/2003.13343}{{\ttfamily 2003.13343}}].

\bibitem{Mack:2009mi}
G.~Mack, \emph{{D-independent representation of Conformal Field Theories in D
  dimensions via transformation to auxiliary Dual Resonance Models. Scalar
  amplitudes}},  \href{https://arxiv.org/abs/0907.2407}{{\ttfamily 0907.2407}}.

\bibitem{Santagata:2022hga}
M.~Santagata, \emph{{Holographic correlators and their (hidden) symmetries}},
  Ph.D. thesis, Southampton U., 2022.

\bibitem{Binder:2019jwn}
D.~J. Binder, S.~M. Chester, S.~S. Pufu and Y.~Wang, \emph{{$ \mathcal{N} $ = 4
  Super-Yang-Mills correlators at strong coupling from string theory and
  localization}}, \href{https://doi.org/10.1007/JHEP12(2019)119}{\emph{JHEP}
  {\bfseries 12} (2019) 119}
  [\href{https://arxiv.org/abs/1902.06263}{{\ttfamily 1902.06263}}].

\bibitem{Chester:2019pvm}
S.~M. Chester, \emph{{Genus-2 holographic correlator on AdS$_{5} \times S^{5}$
  from localization}},
  \href{https://doi.org/10.1007/JHEP04(2020)193}{\emph{JHEP} {\bfseries 04}
  (2020) 193} [\href{https://arxiv.org/abs/1908.05247}{{\ttfamily
  1908.05247}}].

\bibitem{Chester:2020dja}
S.~M. Chester and S.~S. Pufu, \emph{{Far beyond the planar limit in
  strongly-coupled $ \mathcal{N} $ = 4 SYM}},
  \href{https://doi.org/10.1007/JHEP01(2021)103}{\emph{JHEP} {\bfseries 01}
  (2021) 103} [\href{https://arxiv.org/abs/2003.08412}{{\ttfamily
  2003.08412}}].

\bibitem{Chester:2020vyz}
S.~M. Chester, M.~B. Green, S.~S. Pufu, Y.~Wang and C.~Wen, \emph{{New modular
  invariants in $ \mathcal{N} $ = 4 Super-Yang-Mills theory}},
  \href{https://doi.org/10.1007/JHEP04(2021)212}{\emph{JHEP} {\bfseries 04}
  (2021) 212} [\href{https://arxiv.org/abs/2008.02713}{{\ttfamily
  2008.02713}}].

\bibitem{Aprile:2021mvq}
F.~Aprile and M.~Santagata, \emph{{Two particle spectrum of tensor multiplets
  coupled to AdS$_3 \times$S$^3$ gravity}},
  \href{https://doi.org/10.1103/PhysRevD.104.126022}{\emph{Phys. Rev. D}
  {\bfseries 104} (2021) 126022}
  [\href{https://arxiv.org/abs/2104.00036}{{\ttfamily 2104.00036}}].

\bibitem{Bern:2010ue}
Z.~Bern, J.~J.~M. Carrasco and H.~Johansson, \emph{{Perturbative Quantum
  Gravity as a Double Copy of Gauge Theory}},
  \href{https://doi.org/10.1103/PhysRevLett.105.061602}{\emph{Phys. Rev. Lett.}
  {\bfseries 105} (2010) 061602}
  [\href{https://arxiv.org/abs/1004.0476}{{\ttfamily 1004.0476}}].

\bibitem{Zhou:2021gnu}
X.~Zhou, \emph{{Double Copy Relation in AdS Space}},
  \href{https://doi.org/10.1103/PhysRevLett.127.141601}{\emph{Phys. Rev. Lett.}
  {\bfseries 127} (2021) 141601}
  [\href{https://arxiv.org/abs/2106.07651}{{\ttfamily 2106.07651}}].

\bibitem{Farrow:2018yni}
J.~A. Farrow, A.~E. Lipstein and P.~McFadden, \emph{{Double copy structure of
  CFT correlators}}, \href{https://doi.org/10.1007/JHEP02(2019)130}{\emph{JHEP}
  {\bfseries 02} (2019) 130}
  [\href{https://arxiv.org/abs/1812.11129}{{\ttfamily 1812.11129}}].

\bibitem{Lipstein:2019mpu}
A.~E. Lipstein and P.~McFadden, \emph{{Double copy structure and the flat space
  limit of conformal correlators in even dimensions}},
  \href{https://doi.org/10.1103/PhysRevD.101.125006}{\emph{Phys. Rev. D}
  {\bfseries 101} (2020) 125006}
  [\href{https://arxiv.org/abs/1912.10046}{{\ttfamily 1912.10046}}].

\bibitem{Armstrong:2020woi}
C.~Armstrong, A.~E. Lipstein and J.~Mei, \emph{{Color/kinematics duality in
  AdS$_{4}$}}, \href{https://doi.org/10.1007/JHEP02(2021)194}{\emph{JHEP}
  {\bfseries 02} (2021) 194}
  [\href{https://arxiv.org/abs/2012.02059}{{\ttfamily 2012.02059}}].

\bibitem{Albayrak:2020fyp}
S.~Albayrak, S.~Kharel and D.~Meltzer, \emph{{On duality of color and
  kinematics in (A)dS momentum space}},
  \href{https://doi.org/10.1007/JHEP03(2021)249}{\emph{JHEP} {\bfseries 03}
  (2021) 249} [\href{https://arxiv.org/abs/2012.10460}{{\ttfamily
  2012.10460}}].

\bibitem{Cheung:2022pdk}
C.~Cheung, J.~Parra-Martinez and A.~Sivaramakrishnan, \emph{{On-shell
  correlators and color-kinematics duality in curved symmetric spacetimes}},
  \href{https://doi.org/10.1007/JHEP05(2022)027}{\emph{JHEP} {\bfseries 05}
  (2022) 027} [\href{https://arxiv.org/abs/2201.05147}{{\ttfamily
  2201.05147}}].

\bibitem{Herderschee:2022ntr}
A.~Herderschee, R.~Roiban and F.~Teng, \emph{{On the differential
  representation and color-kinematics duality of AdS boundary correlators}},
  \href{https://doi.org/10.1007/JHEP05(2022)026}{\emph{JHEP} {\bfseries 05}
  (2022) 026} [\href{https://arxiv.org/abs/2201.05067}{{\ttfamily
  2201.05067}}].

\bibitem{Lipstein:2023pih}
A.~Lipstein and S.~Nagy, \emph{{Self-dual gravity and color/kinematics duality
  in AdS$_4$}},  \href{https://arxiv.org/abs/2304.07141}{{\ttfamily
  2304.07141}}.

\bibitem{Armstrong:2023phb}
C.~Armstrong, H.~Goodhew, A.~Lipstein and J.~Mei, \emph{{Graviton Trispectrum
  from Gluons}},  \href{https://arxiv.org/abs/2304.07206}{{\ttfamily
  2304.07206}}.

\end{thebibliography}\endgroup

\end{document}